\documentclass[aip
 amsmath,amssymb,
 reprint,%
aps,
]{revtex4-1}

\usepackage{graphicx}
\usepackage{dcolumn}
\usepackage{bm}

\usepackage[utf8]{inputenc}
\usepackage[T1]{fontenc}
\usepackage{mathptmx}
\usepackage[dvipsnames]{xcolor}
\usepackage{comment}

\begin{document}

\preprint{AIP/123-QED}

\title{Wave Steepening and Shock Formation in Ultracold Neutral Plasmas}

\author{M. K. Warrens}
 \affiliation{Rice University, 6100 Main St. Houston, TX, 77005}


\author{N. P. Inman}
\affiliation{Rice University, 6100 Main St. Houston, TX, 77005}
     
\author{G. M. Gorman}
\affiliation{Sandia National Laboratories, 1515 Eubank Blvd SE, Albuquerque, New Mexico 87185, USA}

\author{B. T. Husick}
\affiliation{Rice University, 6100 Main St. Houston, TX, 77005}

\author{S. J. Bradshaw}
\affiliation{Rice University, 6100 Main St. Houston, TX, 77005}

\author{T. C. Killian}
\affiliation{Rice University, 6100 Main St. Houston, TX, 77005}

\date{\today}

\begin{abstract}
We present observations of wave steepening and signatures of shock formation during expansion of ultracold neutral plasmas formed with an initial density distribution that is centrally peaked and decays exponentially with distance. The plasma acceleration and velocity decrease at large distance from the plasma center, leading to central ions overtaking ions in the outer regions and  the development of a steepening front that is narrow compared to the size of the plasma. The  density and velocity  change dramatically across the front, and significant heating of the ions is observed in the region of steepest gradients. For a reasonable estimate of electron temperature, the relative velocity of ions on either side of the front modestly exceeds the local sound speed (Mach number $M \gtrsim 1$). This indicates that by sculpting steep density gradients, it is possible to create the conditions for shock formation, or very close to it, opening a new avenue of research for ultracold neutral plasmas.

\end{abstract}

\maketitle

\section{\label{sec:intro}Introduction}

A shockwave in a plasma is an abrupt jump in pressure, temperature, and density across a narrow front that propagates through the plasma. \cite{ll87} Shocks can form when a driver moves through a plasma 
or a population of particles moves through a background plasma \cite{btr13} at relative speeds that exceed the speed of sound. Ordinary waves can develop into shockwaves through wave steepening, in which nonlinear propagation effects lead to a steepening of leading density gradients. \cite{whi99} 
Plasma shockwaves are observed in many settings, such as laser-induced plasmas, \cite{clp19} supernova remnants, \cite{tre09} and the solar wind. \cite{sbl22}


In this paper, we present experimental observations of wave steepening and signatures of shock formation during the expansion of ultracold neutral plasmas (UNPs). UNPs are created by photoionizing laser cooled atoms \cite{kkb99,kpp07,kgs21} or cold molecules in a supersonic expansion \cite{mrk08}  near the ionization threshold. They have been used to study many different plasma phenomena, such as plasma waves, \cite{kkb00,bsp03,cmk10,mcs13} streaming plasmas, \cite{mcb15} three-body recombination, \cite{klk01,fzr07} instabilities, \cite{zfr08instability} and equilibration  and transport in the strongly coupled regime. \cite{mur01,bcm12,sdm19}  The expansion into surrounding vacuum of a UNP with an initial Gaussian density distribution  has been studied extensively, \cite{kkb00,rha03,ppr04PRA,ppr04PRL,cdd05,gls07,lgs07,msl15} including expansion in a  uniform magnetic field. \cite{zfr08Bfield,sbs22} Recently, expansion of UNPs with an initial density distribution that decays exponentially with distance from the plasma center was characterized experimentally in a field-free configuration \cite{wgb21} and in a quadrupole  magnetic field. \cite{gwb21} Molecular dynamics simulations were presented for each case. \cite{vbz21,bvz23} 

Since UNPs were first created,\cite{kkb99} the development of shockwaves in this extreme environment has been of interest. Early fluid \cite{rha03} and hybrid molecular dynamics \cite{ppr04PRA} models indicated shockwave formation in the low-density outer regions of expanding plasmas with an initial Gaussian density distribution. Experiments failed to detect such features within the limit of experimental sensitivity. \cite{lgs07} It has also been predicted that shocks develop in UNPs with non-Maxwellian electrons.\cite{smd11} More recent molecular dynamics simulations of a Gaussian, quasi-neutral ultracold plasma showed that significant escape of electrons from the outer regions of the plasma and resulting non-neutrality can give rise to shock-like features. \cite{vbz21}  

Early numerical work by Sack and Schamel \cite{ssc85} in a very different context than UNPs showed that expansion into vacuum of a plasma with  an initial step-function density distribution yields wave-steepening and formation of a sharp density peak and shock structure. Following this line of reasoning, the possibility of forming shocks in UNPs sculpted to have initial density distributions with steep density features deviating from an ideal Gaussian has been discussed in multiple contexts.\cite{cmk10,mcs13,mcb15,dmu20} Experimental realizations produced streaming populations of UNPs, displaying a crossover from hydrodynamic \cite{cmk10,kmo12,mcs13} to kinetic behavior \cite{mcb15} with increasing initial density gradients and resulting relative velocities, but signatures of shock formation were not observed. A hydrodynamic model of colliding ultracold neutral molecular plasmas also predicted the formation of shocks.  \cite{mgr15} While not in the UNP context, shockwaves have been seen in the Coulomb explosion of a cold, pure-ion, non-neutral plasma created by photo-excitation well above the ionization threshold such that electrons escape the system.  \cite{vmr20}  

We observe that the shape of the initial density distribution of a UNP largely determines the character of the plasma expansion when magnetic fields are negligible. \cite{wgb21} Plasmas with a Gaussian-shaped initial density distribution, which are well-studied experimentally, expand self-similarly \cite{lgs07} and do not display detectable signatures of wave-steepening and shock formation. As shown in this study, however, plasmas formed with a more sharply peaked distribution that decays exponentially in space \cite{wgb21,gwb21} yield more complex dynamics, and after significant expansion they display many of the traits that characterize the formation of shocks.  

\section{Experimental Methods}
\begin{figure}[ht]
    \centering
   \includegraphics[width = 0.45\textwidth]{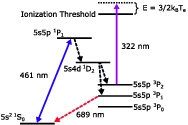}
   \caption{Levels and  transitions for  laser cooling of strontium atoms, trapping in a purely magnetic trap, and subsequent photoionization to form a UNP.}
   \label{fig:atomlevels}
\end{figure}

We create UNPs from strontium atoms that are first laser cooled and trapped in a magneto-optical trap (MOT)\cite{mvs99} operating on the  $^1S_0$-$^1P_1$ principal transition at 461 nm. The MOT uses a quadrupole magnetic field produced by two coils in an anti-Helmholtz configuration.  Fig. \ref{fig:atomlevels} depicts the strontium atom level diagram and relevant decay paths. 1 in $2\times 10^4$ atoms excited to the $^1P_1$ state by the cooling laser will decay into the $^1D_2$ state,\cite{ccm18} after which the atom will return to the ground $^1S_0$ state via the $^3P_1$ state, or decay to the  metastable $^3P_2$ state, with a 2:1 ratio. \cite{xlh03} A fraction of the atoms that decay into the  $^3P_2$ state are trapped in a purely magnetic trap formed by the interaction of the atom's electronic spin with the quadrupole field of the MOT.\cite{nsl03} The resulting trapping potential yields an atomic density distribution that exponentially decays from the center of the trap, which is inherited by the plasma and is given by 
\begin{equation}
n(\vec{r}) = n_0\mathrm{exp}\Bigg [ -\frac{\sqrt{x^2+ ({y^2} + {z^2})/{4}\eta^2}}{\alpha} \Bigg ],
\label{eq:cuspyDensityDensityDistribution}
\end{equation}
where $\alpha = 3k_BT/(8 \mu_B \partial B/\partial x)$ for the magnetic moment $\mu_B$, gradient of the field along the symmetry axis $\partial B/\partial x$, Boltzmann constant $k_B$, and atom temperature $T$. $\eta=1$ for an equilibrated atomic sample, but we observe that the plasma density is best fit with $\eta =0.79$, indicating a lack of thermal equilibrium in the atoms that does not affect the conclusion of the present studies. In this experiment, $\alpha = 0.47$ mm.

After the magnetic trap is loaded with about $10^8$ atoms, the cooling laser beams and the magnetic field are turned off. A 10 ns pulse of 322 nm light photoionizes the cloud of strontium atoms forming a plasma. The electron temperature is set by how far above the ionization threshold the 322-nm beam is detuned \cite{kpp07} as illustrated in Fig. \ref{fig:atomlevels}. Experiments reported here were performed at electron temperatures ($T_e = 160$ K) high enough to avoid the effects of three body recombination, which increases at lower temperatures. \cite{kkb00} 


The early stages of UNP equilibration are described in Ref. \onlinecite{kpp07}. After photionization, electrons equilibrate in the first few hundred ns. On the timescale of the ion plasma oscillation of about 1 $\mu$s, the ions undergo disorder-induced heating (DIH) \cite{csl04,lsm16} to a temperature of $T_i\approx e^2/(12\pi\epsilon_0 a k_B)\approx 1$\,K  depending on the density, where $e$ is the elementary charge, $\epsilon_0$ is the vacuum permittivity, $k_B$ is the Boltzmann constant, and $a=[3/(4\pi n_i)]^{1/3}$ is the Wigner-Seitz radius and $n_i$ is the ion density. Following intraspecies equilibration, the plasma expands into the surrounding vacuum.

At an adjustable time after plasma creation, the Sr$^+$ ions are imaged using laser-induced fluorescence (LIF) on the $^2S_{1/2}$-$^2P_{1/2}$ transition at 422 nm. \cite{cgk08} Ions are excited with a thin (1 mm) sheet of laser light illuminating the central plane of the plasma. LIF  emitted perpendicular to the light sheet is imaged onto an intensified CCD camera using a 1:1 optical relay with a resolution of  0.1 mm.

\section{Results}

 \subsection{Plasma Expansion and Wave Steepening}

\begin{figure}[!ht]
    \centering
   \includegraphics[width = 0.48\textwidth]{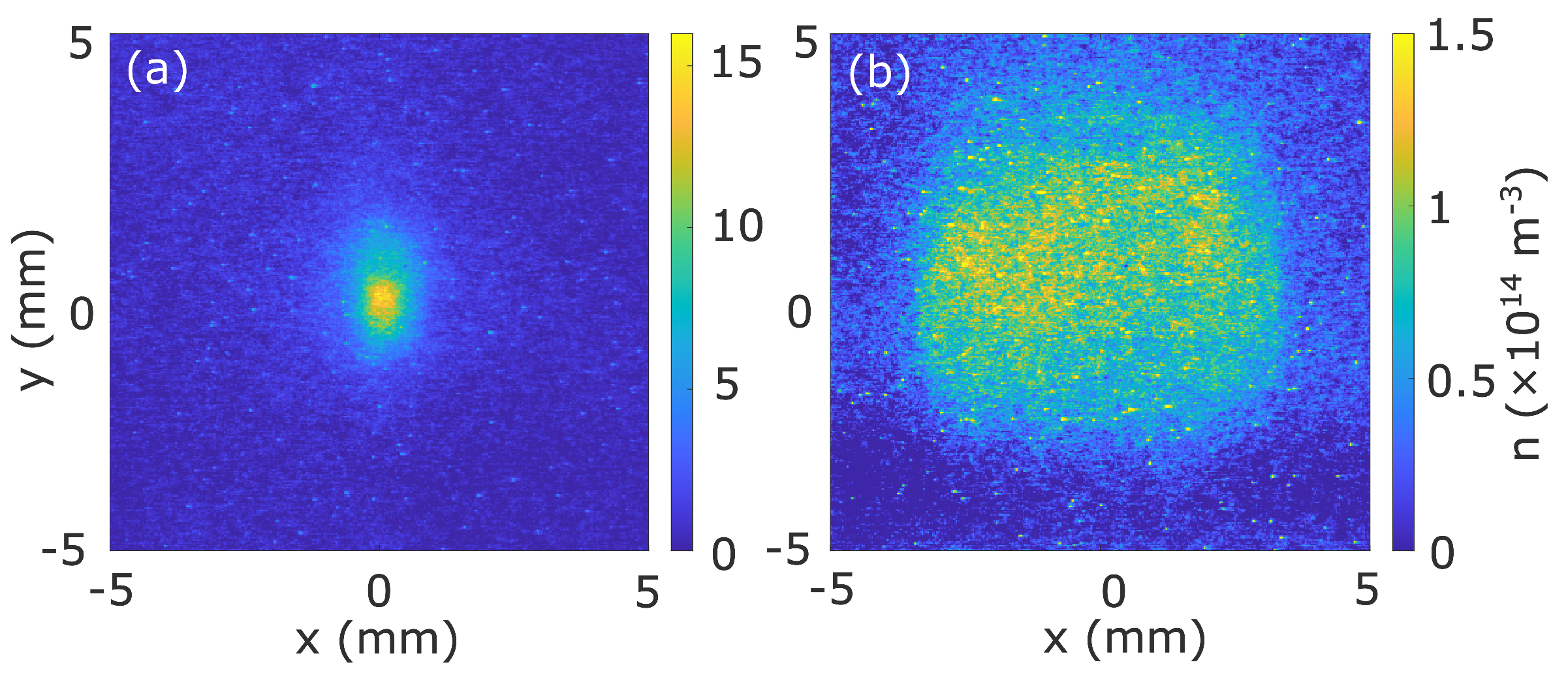}
   \caption{False color images of the density in a cut through the plasma center. (a) 0.3 $\mu$s after plasma creation. Tighter confinement along the $x$-axis of the atoms and the resulting plasma creates a larger density gradient and hydrodynamic pressure driving expansion along that axis. (b) 12 $\mu$s after plasma creation. The aspect ratio of the plasma has inverted, and sharp transitions in the density are visible at the leading edge of expansion along the $x$-axis, suggesting wave steepening and shock formation. }
   \label{fig:plasmaImages}
\end{figure}

Figure \ref{fig:plasmaImages}(a) shows a false-color image of the density profile of a UNP 0.3 $\mu $s after plasma formation, which is close to the initial plasma density distribution. The image is of the plane illuminated by the LIF laser, passing through the center of the plasma along the axis of cylindrical symmetry of the initial atomic density distribution.

UNP expansion is driven predominantly by the electron thermal pressure gradient. A hydrodynamic  description yields the local acceleration given by\cite{lgs07}  
\begin{equation}
    \label{eq:hydroAccel}
    \dot{\vec{u}} = -\frac{k_BT_e}{m_i}\frac{\nabla n}{n},
\end{equation}
where $u$ is the bulk plasma velocity,  $m_i$ is the mass of a strontium ion, $T_e$ is the electron temperature, and $n$ is the number density. This expression assumes quasi-neutrality,  $n_i = n_e = n$, which is typically a good approximation for UNPs and holds when the electron temperature is low enough for the Debye length to be much less than the spatial scale for density variation in the plasma. \cite{kpp07} The Debye length for the peak density ($n=15\times 10^{14}\,\rm{m}^{-3}$ and $T_e=160$\,K) is 23 $\mu$m.

 The steepest plasma gradients are along the $x$-axis (see Eq.\ \ref{eq:cuspyDensityDensityDistribution}), producing the highest plasma acceleration along this axis and an inversion of the aspect ratio during the expansion as shown in the image taken at 12\,$\mu$s in Fig.\ \ref{fig:plasmaImages}(b). Sharp transitions in the density are visible at the edges of the plasma experiencing the fastest expansion. We focus on the evolution of the density along the $x$-axis to look for signatures of wave steepening and shock formation. 

\begin{figure}[!ht]
    \centering
    \includegraphics[width = 0.45\textwidth]{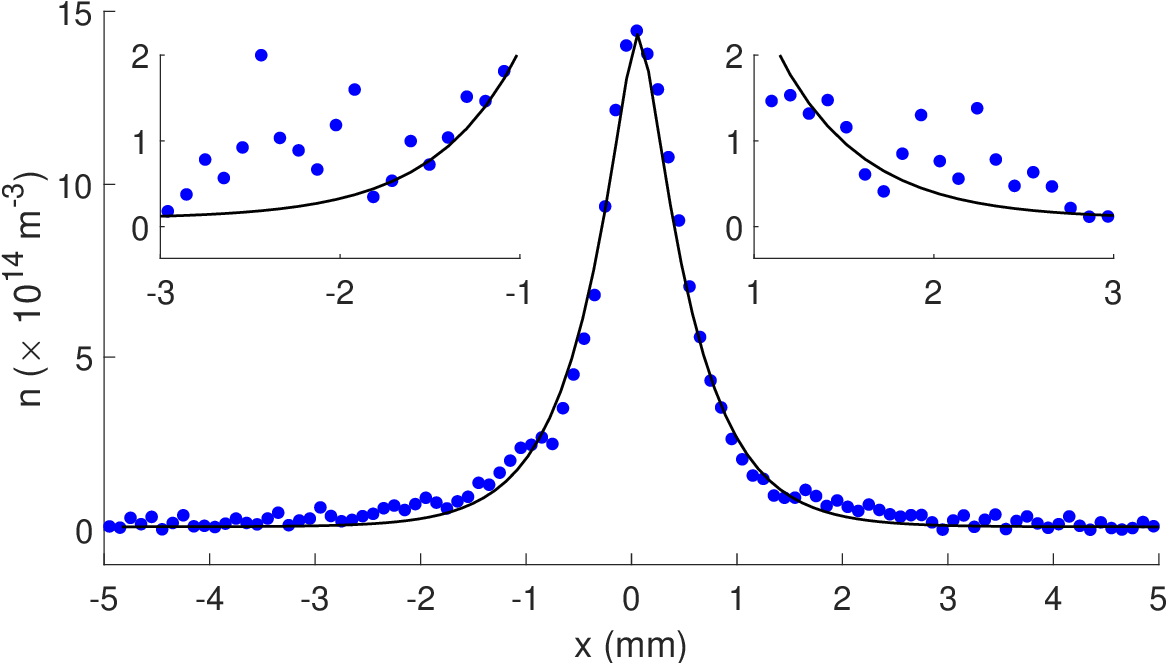}
    \caption{Initial density along the $x$-axis at $y\approx z \approx 0$  from the image in Fig.\ \ref{fig:plasmaImages}(a) and a transect of a 2D exponential fit.}
    \label{fig:initialImage}
\end{figure}

A transect of the initial density at $y \approx 0$ is shown in Fig. \ref{fig:initialImage}. To fit the data, the 2D exponential (Eq. 1) is averaged
over a width in the $z$-axis corresponding to the thickness of the LIF light sheet. This produces a slight rounding of the
peak. The overall match to the data is good,  
although, there is some excess plasma in the regions of approximately $1<|x|<3$ mm. This pedestal  may contribute to wave steeping as it creates a region of lower
$\nabla n/n$ and therefore lower initial acceleration velocity (Eq. \ref{eq:hydroAccel}). 

For an exponential density distribution, the hydrodynamic expression for acceleration (Eq. \ref{eq:hydroAccel}) yields a step function,  uniform on either side of the plasma, but with opposite signs. This would initially give rise to a velocity profile that is also a step function.
The evolution is more complex than this simple picture because of initial deviations from a perfect exponential and the density distribution quickly evolves away from an exponential.

\begin{figure}[!ht]
    \centering
    \includegraphics[width = 0.45\textwidth]{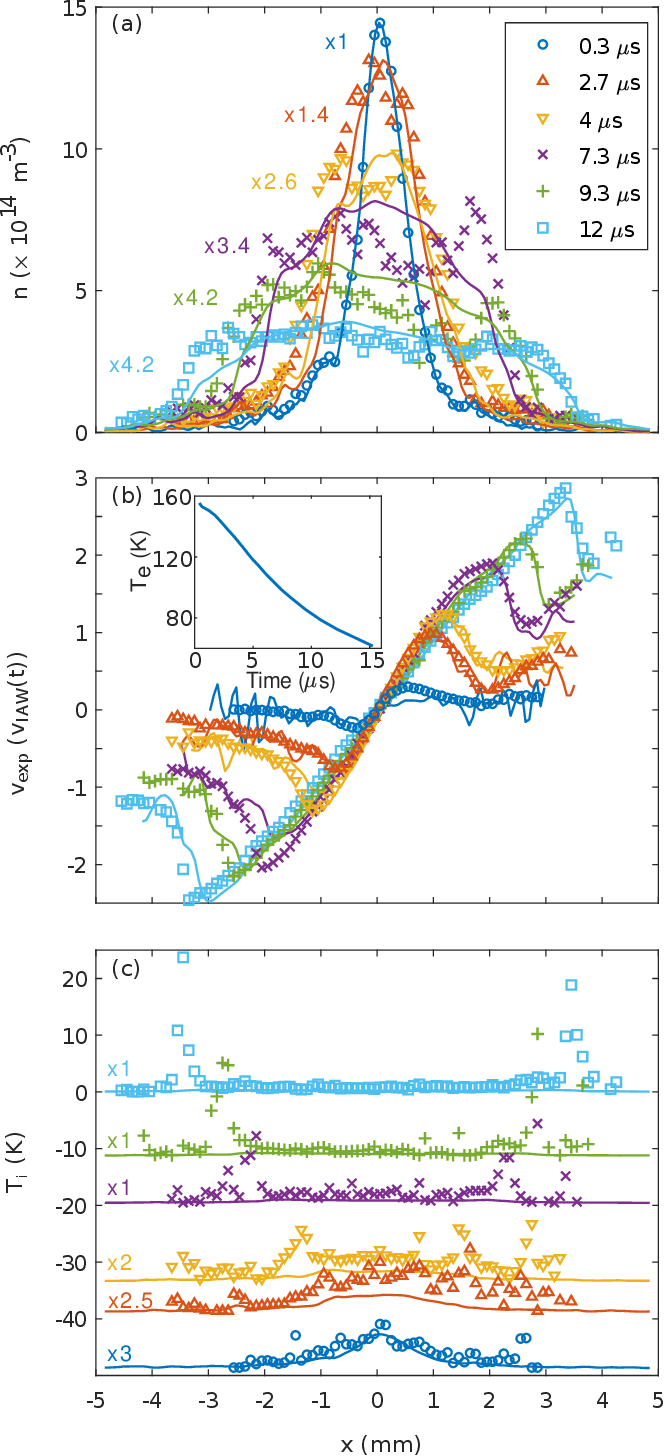}
    \caption{ 
   Wave steepening during plasma expansion. Symbols are experimental data and solid lines are results of a fluid simulation described in the text. (a) Density transects along $x$-axis for $y\approx z \approx 0$.  (b) Velocity transects scaled by the ion acoustic wave speed ($v_{IAW}$ = 128, 120, 115, 101, 94, and 86 m/s in order of increasing time). (c) Ion temperature transects for different times in the plasma expansion.  $T_i$ is offset by -48.8, -38.8, -33.3, -19.6, -11.3, and 0 K in order of increasing time. The offset is for clarity and is proportional to time to show the front propagation velocity. The calculated electron temperature, $T_e$ (K), with respect to time is included as an inset in (b). Data and simulations are multiplied by the factors indicated in the figure to improve visibility.} 
    \label{fig:anatomyOfAShockWithSymbols}
\end{figure}

Figure \ref{fig:anatomyOfAShockWithSymbols} shows transects of the experimentally measured plasma density (a), bulk velocity (b), and ion temperature (c) along the $x$-axis at various times after photoionization, compared to fluid simulations produced by SPRUCE, a multipurpose fluid code under development in the Solar Physics Research Group at Rice University.\cite{Szypko2023Modeling}

The density (Fig.\ \ref{fig:anatomyOfAShockWithSymbols}(a)) becomes flat in the center within the first few microseconds of evolution. As time progresses, the size of the central region of flat plasma density increases as the overall plasma expands and the peak density decreases. The leading edge of the plasma expansion accelerates with time as electron thermal energy is converted into directed kinetic energy of the ion expansion. \cite{kkb00}

The expansion for the exponential plasma is clearly not self-similar. It is instructive to contrast it with the
expansion of a spherical Gaussian density distribution \cite{lgs07,wgb21} given by $n=n_0\rm{exp}(-r^2/2\sigma_0^2)$, for which the hydrodynamic force (Eq.\ \ref{eq:hydroAccel}) yields an acceleration that increases linearly from zero at the plasma center. This yields a hydrodynamic velocity of ${\bf{v}}=\gamma \textnormal{(t)}\bf{r}$ that increases linearly with distance from the plasma center and initially increases with time, resulting in a self-similar expansion, $\sigma_0\rightarrow \sigma(t)$. Both $\gamma(t)$ and $\sigma(t)$ can be expressed analytically. \cite{lgs07,kpp07} 

 For the exponential plasma, the acceleration is initially much larger near the center than for a Gaussian plasma of comparable size and electron temperature, but the acceleration of the Gaussian plasma will be greater at large radial distance. An exponential density is therefore susceptible to the development of a velocity distribution that displays extrema away from the plasma center, as shown in Fig.\ \ref{fig:anatomyOfAShockWithSymbols}(b).
For the earliest time point, the sharply peaked center of the plasma may experience significant non-neutrality in a region on the order of the Debye length, which is 23 $\mu$m for the peak density. This may create a local Coulomb explosion 
that also contributes to the acceleration of ions from the central region. \cite{wgb21}  


The bulk velocity in the flat central region of the exponential plasma is well-approximated by the linear relationship $v=x/t$ after the first few microseconds, indicating that plasma in this region quickly evolves to a state of expansion with vanishing acceleration, as expected for a flat density (Eq.\ \ref{eq:hydroAccel}). Note that Fig.\ \ref{fig:anatomyOfAShockWithSymbols}(b) shows the velocity scaled by the ion acoustic wave velocity, which will be discussed in detail below and is given for each trace in the caption of the figure.  
The largest observed peak expansion velocities are $\sim 200$\,m/s at 7\,$\mu$s. 

The peaks of the ion velocity for positive and negative $x$ precisely correspond to the edges of the central density plateau. In the velocity profiles, it is evident that from $0.3-7.3\,\mu$s, higher-velocity plasma is overtaking lower-velocity plasma in front of it yielding wave steepening. 
By $\sim 7\,\mu$s this creates well-defined high-density, high-velocity fronts that propagate through the low-density, low-velocity plasma in front of them, with a large jump in density and velocity across the front.

\begin{figure}[!ht]
    \centering
    \includegraphics[width = 0.45\textwidth]{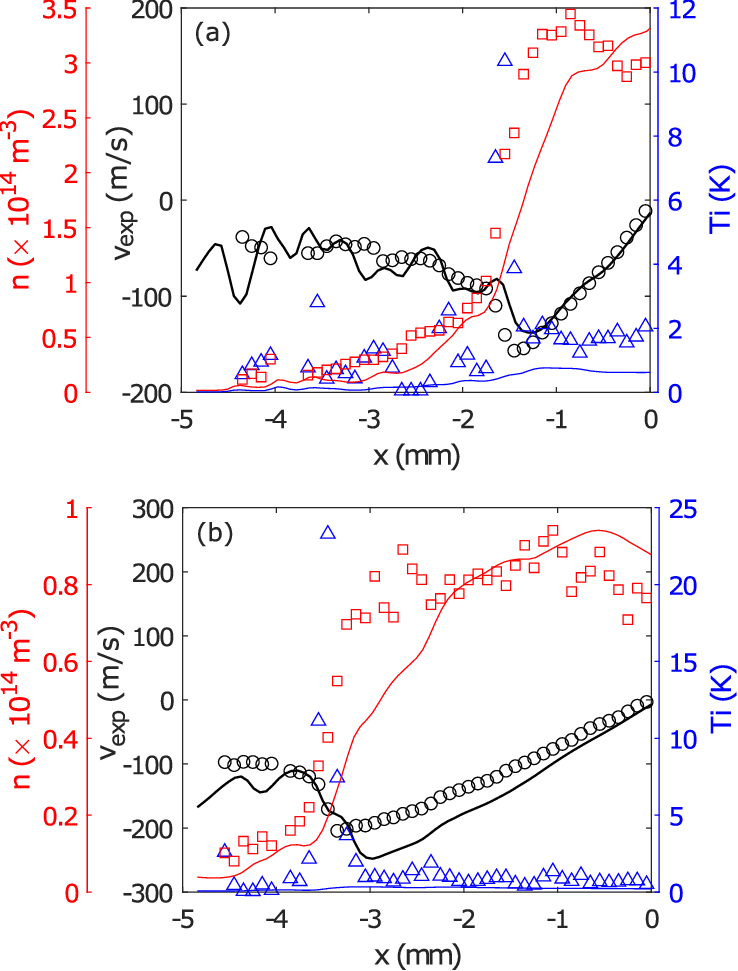}
    \caption{Transects for  the plasma  expansion velocity ($\circ$), density ($\square$), and ion temperature ($\triangle$) for the left half of the plasma at (a) 5.3 $\mu s$ and (b) 12.0 $\mu s$ expansion times. The principal features in each trace overlap spatially. Solid lines are numerical simulations.}
    \label{fig:TripleTransect}
\end{figure}

At $12\,\mu$s the velocity peaks and gradients are extremely sharp, with a length scale for the front (from 75\%  to 25\% of peak density) of 0.4 mm, much smaller than the several-millimeter scale of the overall plasma. 
The jump in velocity is over 100\,m/s.   The spatial overlap of the sharp density and velocity features is clearly seen in the expanded views shown in Fig. \ref{fig:TripleTransect}.

The experimental and simulation results agree reasonably well. The simulation treats the plasma as an ideal quasi-neutral fluid in two dimensions with one continuity and momentum equation, and separate electron and ion energy equations. The fluid equations are solved on a uniform Cartesian grid using Barton's method for monotonic transport to compute transport derivatives, second-order central finite differences for all other derivatives, and a second-order Runge-Kutta time integration scheme.\cite{Szypko2023Modeling}  Barton's method is a simple, yet robust, method of differentiation for discontinuous fluid problems that result in shock conditions without post-shock oscillations present in other methods.\cite{Centrella1984} Open boundary conditions are implemented by choosing a large simulation domain $|x|,|y|<1.2$\,cm and limiting spatial derivatives at the boundary through the use of two ghost cells, which copy the fluid characteristics of the nearest interior cell. 

The plasma density is initialized on the domain by interpolating the density distribution shown in Fig.\ \ref{fig:plasmaImages} 
for distance from the center $|r|<5$\,cm and filling all cells exterior to this radius with an exponential fit to the distribution in Fig.\ \ref{fig:plasmaImages} superimposed on a small uniform background ($10^{-4}$ of peak density) to avoid simulation instability at vanishing density. The initial electron temperature is set to $160$\,K to match the energy expected from the detuning of the ionizing laser above resonance. The ion temperature profile is set to the initial measured value, after disorder-induced heating.

Electron thermal energy redistributes rapidly across the plasma because of the large electron mean free path (MFP) ($\lambda_{e}\gg\alpha$). To capture this behavior, the electrons are held in global thermal equilibrium by replacing the electron temperature distribution with a uniform value $\bar{T_e}$ after each time step:
\begin{equation}
\label{eq:global_temp}
    \bar{T_e}=\frac{1}{N}\int{nT_e(x,y)\,dxdy},
\end{equation}
where $N=\int n\,dxdy$. Solving the electron energy equation and redistributing the thermal energy in this way allows for the capture of expansion-induced cooling while conserving thermal energy and avoiding unphysical temperature gradients.

The validity of this quasi-kinetic treatment of the electrons, where the electrons are globally hydrodynamic but locally kinetic, has been demonstrated for UNPs with a variety of initial density distributions such as a spherically symmetric Gaussian density distribution imprinted with planar ion holes and periodic perturbations that excite ion acoustic waves.\cite{Szypko2023Modeling} The simulations in two dimensions accurately capture the key elements of the plasma expansion dynamics. The pressure gradients that drive plasma acceleration are identical in each case as is the resulting expansion-induced cooling for thermal energy density, $e=P/(\gamma-1)$, where $P$ is the plasma pressure, which is dominated by the electrons. $\gamma=2$ is the two-dimensional adiabatic index used in the simulations. Likewise, the plasma sound speed is identical in each case due to its independence of the adiabatic index under locally kinetic conditions where $T_e\gg T_i$. 

One consequence of simulating in two dimensions is the plasma is effectively treated as uniform in the $z$ direction. Thus, the simulations will neglect ion flux out of the central plane, causing the absolute density to be higher in the simulations. The lack of self-similarity in plasma expansion can result in differences in the evolution of the density distribution. However, the absolute density does not play a role in driving the plasma expansion or the sound speed. The agreement between experiment and simulation for the renormalized density profile and absolute velocity distribution gives confidence that the electron temperature predicted by the simulation is a reasonable estimate of the actual electron temperature. We use this electron temperature to calculate the ion acoustic wave speed for Fig.\ \ref{fig:anatomyOfAShockWithSymbols}(b).

\subsection{Mach Number}

In addition to jumps in velocity and density over a small length scale, a shockwave displays a relative flow velocity that exceeds the local sound speed so that information of the oncoming shock cannot propagate before it.
This is characterized with a Mach number $M>1$. We define $M$ for our system as
\begin{equation}
    M = \frac{v_{rel}}{v_{s}}, 
    \label{eq:Mach}
\end{equation}
where $v_{rel}$ is the relative velocity between the plasma populations on either side of the front, and $v_s$ is the sound speed in the plasma. For an unmagnetized plasma, the sound speed is the ion acoustic wave speed for kinetic electrons and hydrodynamic ions which is given by

\begin{equation}
    \label{eq:vIAW}
    v_{IAW} = \sqrt{\frac{k_B}{m_i}(T_e+\gamma_iT_i)},
\end{equation}
where $k_B$ is the Boltzmann constant, $m_i$ is the ion mass, $T_e$ is the electron temperature,  and $T_i$ is the ion temperature. $\gamma_i$ is the adiabatic index for the ions and is $5/3$ because acoustic wave compression is isentropic.

Knowledge of the electron temperature is essential to estimate the ion acoustic wave speed. We are unable to directly measure the electron temperature, however, the initial electron temperature is set by the ionization laser detuning above threshold \cite{kkb99,kpp07,gls07} ($T_e(0)=160$ K for the data discussed here). We can also reliably assume global thermal equilibrium for the electrons because the MFP is on the order of the size of the plasma and the collision time is short compared to the plasma expansion time. Initially, for 160 K and a density of $15 \times 10^{14}$ m$^{-3}$, the MFP for an electron moving with the thermal velocity is 2.4 mm. 

We compare three different estimates of the electron temperature and ion acoustic wave speed for each time point during the plasma expansion. An upper bound for both is given by $T_e=T_e(0)$, which neglects adiabatic cooling of the electrons during the plasma expansion. This yields a lower limit on the Mach number. Alternatively, an approximate value of $T_e(t)$ can be obtained from a simple estimate of the volume $V$ of the plasma at each time derived from the plasma images assuming cylindrical symmetry and adiabatic cooling of the electrons according to $T_eV^{2/3}=constant$.\cite{ppr04PRA} Finally, from the 2D fluid simulation, we extract the simulated $T_e(t)$, which we take as our most reliable value.
Figure \ref{fig:anatomyOfAShockWithSymbols}(b) shows the velocity transects scaled by the ion acoustic wave speed using $T_e(t)$ from the hydrodynamic simulation. For calculating $v_{IAW}$, the ion temperature is negligible.

\begin{figure}[!ht]
    \centering
   \includegraphics[width = 0.45\textwidth]{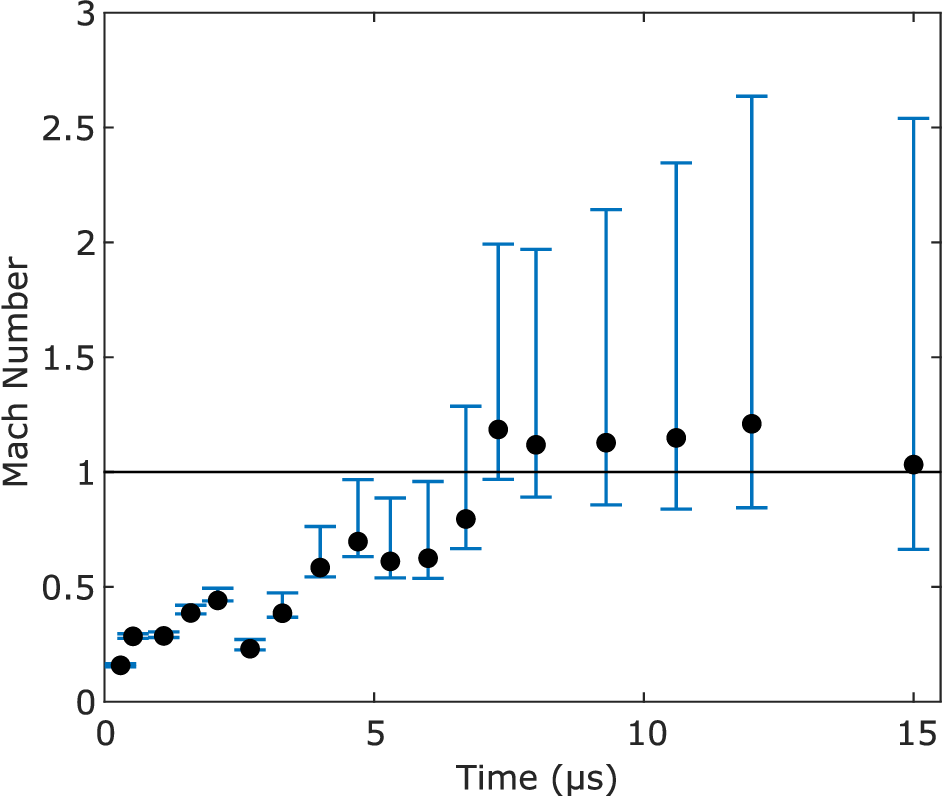}
   \caption{Mach number evolution for the plasma front for $x<0$. Upper and lower bounds are described in the text. }
   \label{fig:MachNumber}
\end{figure}

Figure \ref{fig:MachNumber} shows the Mach number as a function of time for the left side of the plasma ($x<0$).  The relative velocity is taken as the difference between the peak velocity and the mean velocity in the region of relatively constant velocity near the edge of the plasma. Initially, $M$ < 1, but for expansion times greater than  $\sim 7\,\mu$s, the Mach number slightly exceeds unity (or comes very close to unity when using the lower bound). $7.3\,\mu$s is also the time where the velocity profile develops a sharp kink at the peak (Fig.\ref{fig:anatomyOfAShockWithSymbols}(b)), further suggesting the formation of a shockwave.


\subsection{Localized Heating}
The plasma also displays localized heating in the ion temperature where the density and velocity gradients are highest, as shown by Figs. \ref{fig:anatomyOfAShockWithSymbols}(c) and \ref{fig:TripleTransect}. Initially ({\it e.g.} for plasma evolution time $t=0.3\,\mu$s), the ion temperature is relatively smooth. Disorder-induced heating (DIH) \cite{csl04,lsm16} should yield a density-dependent temperature, with the highest value in the plasma center of  1.2 K for density of $14\times10^{14}$ m$^{-3}$.  The observed temperature is  higher (2.6 K in the center). This may arise from kinetic effects and initial non-neutrality in the region of very high density gradient in the plasma center. \cite{wgb21} The center of the plasma adiabatically cools throughout the plasma expansion, as expected. 

As wave steepening develops in the plasma, increases in the ion temperature form and grow at the density and velocity jump. This behavior becomes prominent at 7.3 $\mu$s in Fig. \ref{fig:anatomyOfAShockWithSymbols}(c). Ion temperatures in the front exceed 10 K where the density and velocity jumps are greatest. Plotting the density, velocity, and temperature transects on the same plot, as shown in Fig. \ref{fig:TripleTransect} for 7.3 $\mu$s, shows the overlap of the temperature features with the front.
The numerical simulation only displays a small temperature rise in the region of the front, most clearly seen in Fig.\ \ref{fig:TripleTransect}(a). This suggests that the temperature spikes arise from kinetic effects, which are not included in the simulation. 

It is important to consider that an overestimation in the ion temperature could arise from artificial spectral broadening introduced by the finite resolution of the imaging system and analysis. This is a concern because of the large velocity gradient at the location of the temperature spike, and because we measure the spectra over a finite region size (0.1 mm). 

 The magnitude of the largest velocity gradient in the data is $3.6\times 10^{5}$ $s^{-1}$ at 12 $\mu$s (Fig. \ref{fig:anatomyOfAShockWithSymbols}(b)). For a 0.1 mm region, this introduces a  velocity spread of $\Delta v=36$\,m/s. A simple upper bound on the increase would produce in the temperature extracted from spectral analysis is $\Delta T\approx m\Delta v^2/2k_B=7$\,K, which does not account for the observed temperature in excess of 20\,K. 
 
 To better estimate the the spectral broadening due to the velocity gradient, we developed a simulation that divides each 0.1 mm region into smaller subregions in which the velocity spread can be ignored. The simulation creates spectra for each subregion and then adds them together to create the simulated spectrum for the original 0.1 mm region. Simulated spectra are then analyzed in the same way as the experimental data and yield only about 2\,K of increase in the measured temperature due to the finite region size for the largest velocity gradients.




 


\section{Ion Mean Free Path}
As discussed above, the electron MFP is on the order of the plasma size, and the collision time is short compared to the timescale for plasma expansion. This allows us to assume global thermal equilibrium for the electrons.
The situation for ions is more complex. The MFP for an ion moving at the thermal velocity for 1 K is 28 $\mu$m for a typical density of $1\times 10^{14}$\,m$^{-3}$. This is small relative to the size of the plasma, which is consistent with the lack of global thermal equilibrium observed in the system (Fig.\ \ref{fig:anatomyOfAShockWithSymbols}(c)).  In regions away from the expanding fronts, the ions quickly establish local thermal equilibrium because the collision time and the timescale for DIH is $\sim1\,\mu$s. \cite{lsm16}

At the front, however, the large relative velocities of ions in the low density and high density sides of about 100 m/s yield much longer MFPs. For the typical density on the high density side of  $1\times 10^{14}$\,m$^{-3}$, the MFP is 6 mm, which is very large compared to the size of the shock front. Therefore, we expect a significant fraction of the population of low density ions in front of the shock to penetrate into the high density region overtaking it. A detailed description of this process would require a calculation of the stopping power of the ions, including possible effects from strong coupling, \cite {mro96} which is beyond the scope of this paper. We expect that the population of ions in the high density region expanding outward is in local thermal equilibrium with itself due to the high collisionality. Any streaming population is too small to detect relative to the high density population.  LIF spectra for plasma in the transition region are fit well with a line shape that assumes local thermal equilibrium.

\section{Conclusion}
We have presented evidence of wave steepening and signatures of shock formation in UNPs formed with an initially sharply peaked density distribution. 
As the plasma expands, sharp transitions in density and velocity form at expanding fronts that also display localized ion heating. Using the electron temperature taken from a fluid simulation, Mach numbers slightly greater than unity are observed. This work establishes that conditions for shock formation,  or near to it,  can be obtained in UNPs with sculpted density distributions.

This data presents a rich challenge for testing hydrodynamic and kinetic theories and  for searching for effects of strong coupling on viscosity through damping of the shock features. Future work can explore the variation of the observed features with plasma density and initial electron temperature. Expanding the field of view for plasma imaging would enable exploration of whether wave steepening continues or is eventually balanced by dissipation.




\begin{acknowledgments}
This work was supported by the NSF/DOE Partnership in Basic Plasma Science and Engineering Award Number 2107709, the National Science Foundation Graduate Research Fellowship Program under Grant Number 1842494, and NSF-CAREER award AGS-1450230. Any opinions, findings, and conclusions or recommendations expressed in this material are those of the authors and do not necessarily reflect the views of the National Science Foundation.
\end{acknowledgments}

\section*{Data Availability}
The data that support the finding of this study are available from the corresponding author upon reasonable request.


@Article{bwp01,
  author = {Badziak, J.  and Woryna, E.  and Parys, P.  and Platonov, K. Yu. and Jab\l{}o\ifmmode \acute{n}\else \'{n}\fi{}ski, S.  and Ry\ifmmode \acute{c}\else \'{c}\fi{}, L.  and Vankov, A. B. and Wo\l{}owski, J. },
  title = {Fast Proton Generation from Ultrashort Laser Pulse Interaction with Double-Layer Foil Targets},
  journal = {Phys. Rev. Lett.},
  volume = {87},
  number = {21},
  pages = {215001},
  year = {2001},
  publisher = {American Physical Society}
}

@book{btr13,
  AUTHOR =       {A. Balogh and R. A. Treumann},
  TITLE =        {Physics of Collisionless Shocks},
  PUBLISHER =    {Springer},
  YEAR =         {2013},
  address =      {New York, New York},
}

@article{bcm12,
    author = {Bannasch, G. and Castro, J. and McQuillen, P. and Pohl, T. and Killian, T. C.},
    title = {Velocity Relaxation in a Strongly Coupled Plasma},
    doi = {10.1103/PhysRevLett.109.185008},
    issn = {0031-9007},
    journal = {Phys. Rev. Lett.},
    month = {nov},
    number = {18},
    pages = {185008},
    volume = {109},
    year = {2012}
}

@article{bcd88,
Author = "A.R. Bell and P. Choi and A.E. Dangor and O. Willi and D.A. Bassett",
Journal = "{Phys. Rev. Lett.}",
Year = 1988,
Volume = 38,
Number = 3,
Pages =  1363 
}

@article{bbl19,
author = {Bergeson, S. D. and Baalrud, S. D. and {Leland Ellison}, C. and Grant, E. and Graziani, F. R. and Killian, T. C. and Murillo, M. S. and Roberts, J. L. and Stanton, L. G.},
title = {Exploring the crossover between high-energy-density plasma and ultracold neutral plasma physics},
issn = {10897674},
journal = {Phys. Plasmas},
number = {10},
pages = {100501},
publisher = {AIP Publishing LLC},
volume = {26},
year = {2019}
}

@article{bsp03,
  title = {Neutral-plasma oscillations at zero temperature},
  author = {Bergeson, S. D. and Spencer, R. L.},
  journal = {Phys. Rev. E},
  volume = {67},
  issue = {2},
  pages = {026414},
  numpages = {5},
  year = {2003},
  month = {Feb},
  publisher = {American Physical Society},
  doi = {10.1103/PhysRevE.67.026414},
}

@article{bra13,
author = {Bradshaw, Stephen J. and Raymond, John},
title={Collisional and Radiative Processes in Optically Thin Plasmas},
doi = {10.1007/s11214-013-9970-0},
issn = {0038-6308},
journal = {Space Sci. Rev.},
keywords = {microphysical processes,plasma,solar atmosphere},
mendeley-tags = {plasma,solar atmosphere},
month = {mar},
number = {2-4},
pages = {271},
volume = {178},
year = {2013}
}

@article{ccb05,
author = {F. Cornolti and F. Ceccherini and S. Betti and F. Pegoraro},
title = {Charged state of a spherical plasma in vacuum},
publisher = {APS},
year = {2005},
journal = {Phys. Rev. E},
volume = {71},
number = {5},
eid = {056407},
pages = {056407}
}

@article{cdd05,
author = {E. A. Cummings and J. E. Daily and D. S. Durfee and S. D. Bergeson},
title = {Flourescence Measurements Of Expanding Strongly Coupled Neutral Plasmas},
journal = {Phys. Plasmas},
year = {2005},
volume = {12},
number = {12},
pages = {123501}
}

@article{cga04, 
author = {C. Courtois and R. A. D. Grundy and A. D. Ash and D. M. Chambers and N. C. Woolsey and R. O. Dendy and K. G. McClements}, 
title = {Experiment on Collisionless Plasma Interaction with Applications to Supernova Remnant Physics},
year = {2005}, 
journal = {Phys. Plasmas},
volume = {11}, 
pages = {3386}
}

@article{cgk08,
    Author = {J. Castro and H. Gao and T. C. Killian},
    Journal = {Plasma Phys. Control. Fusion},
    Title = {Using sheet fluorescence to probe ion dynamics in an ultracold neutral plasma},
    Pages = {124011},
    Volume = {50},
    Year = {2008}}

@article{cgt05,
author = {B. J. Claessens and S. B. van der Geer and G. Taban and E. J. D. Vredenbregt and O. J. Luiten},
title = {Ultracold Electron Source},
year = {2005},
journal = {Phys. Rev. Lett.},
volume = {95},
number = {16},
pages = {164801}
}

@Article{ckz00,
  author = {E. L. Clark and K. Krushelnick and M. Zepf and F. N. Beg and M. Tatarakis and A. Machacek and M. I. K. Santala and I. Watts and P. A. Norreys and A. E. Dangor},
  title = {Measurements of Energetic Proton Transport through Magnetized Plasma from Intense Laser Interactions with Solids},
  journal = {Phys. Rev. Lett.},
  volume = {85},
  number = {8},
  pages = {1654},
  numpages = {3},
  year = {2000}
}

@article{cmk10,
Author = "J. Castro and P. McQuillen and T. C. Killian",
Title = { Ion Acoustic Waves in Ultracold Neutral Plasmas},
Journal = "{Phys. Rev. Lett.}",
Year = 2010,
Volume = 105,
Pages = 065004}

@article{csl04,
Author = "Y. C. Chen and C. E. Simien and  S. Laha and P. Gupta  and  Y. N. Martinez and P. G. Mickelson and S. B. Nagel and T. C. Killian",
Title = {Electron Screening and Kinetic-Energy Oscillations in a Strongly Coupled Plasma},
Journal = "{Phys. Rev. Lett.}",
Year = 2004,
Volume = 93,
Pages =  265003}

@article{ccm18,
  title = {Alkaline-Earth Atoms in Optical Tweezers},
  author = {Cooper, Alexandre and Covey, Jacob P. and Madjarov, Ivaylo S. and Porsev, Sergey G. and Safronova, Marianna S. and Endres, Manuel},
  journal = {Phys. Rev. X},
  volume = {8},
  issue = {4},
  pages = {041055},
  numpages = {19},
  year = {2018},
  month = {Dec},
  publisher = {American Physical Society},
  doi = {10.1103/PhysRevX.8.041055},
}


@Article{clp19,
AUTHOR = {B. Campanella and S. Legnaioli and S. Pagnotta and F. Poggialini and V. Palleschi},
TITLE = {Shock Waves in Laser-Induced Plasmas},
JOURNAL = {Atoms},
VOLUME = {7},
YEAR = {2019},
NUMBER = {2},
ARTICLE-NUMBER = {57},
ISSN = {2218-2004},
ABSTRACT = {The production of a plasma by a pulsed laser beam in solids, liquids or gas is often associated with the generation of a strong shock wave, which can be studied and interpreted in the framework of the theory of strong explosion. In this review, we will briefly present a theoretical interpretation of the physical mechanisms of laser-generated shock waves. After that, we will discuss how the study of the dynamics of the laser-induced shock wave can be used for obtaining useful information about the laser–target interaction (for example, the energy delivered by the laser on the target material) or on the physical properties of the target itself (hardness). Finally, we will focus the discussion on how the laser-induced shock wave can be exploited in analytical applications of Laser-Induced Plasmas as, for example, in Double-Pulse Laser-Induced Breakdown Spectroscopy experiments.}
}


@article{dai02,
author = {H. Daido},
journal = {Rep. Prog. Phys.},
pages = {1513},
volume = {65},
year = {2002}
}

@article{dra09,
author = {R. P. Drake},
journal = {Phys. Plasmas},
pages = {055501},
volume = {16},
year = {2009}
}

@article{dmu20,
  author = {Dharodi, Vikram S. and Murillo, Michael S.},
title = {Sculpted ultracold neutral plasmas},
  journal = {Phys. Rev. E},
  volume = {101},
  issue = {2},
  pages = {023207},
  numpages = {14},
  year = {2020},
  month = {Feb},
  publisher = {American Physical Society}
}


@article{ehm15contPlasmaPhys,
Author = "S. A. Ema and M. R. Hossen and A. A. Mamun",
Journal = "{Contrib. Plasma Phys.}",
Year = 2015,
Volume = 55,
Number = 8,
Pages =  596 
}

@article{ehm15PhysPlasmas,
Author = "S. A. Ema and M. R. Hossen and A. A. Mamun",
Title = {Linear and nonlinear heavy ion-acoustic waves in a strongly coupled plasma},
Journal = "{Phys. Plasmas}",
Year = 2015,
Volume = 22,
Pages =  092108
}

@article{fkv18,
author = {Franssen, J. G.H. and Kromwijk, J. M. and Vredenbregt, E. J.D. and Luiten, O. J.},
issn = {13616455},
journal = {J. Phys. B},
number = {3},
pages = {035007},
publisher = {IOP Publishing},
title = {{Energy spread of ultracold electron bunches extracted from a laser cooled gas}},
volume = {51},
year = {2018}
}

@article{fsk05,
Author = "J. Fuchs and Y. Sentoku and S. Karsch and J. Cobble and P. Audebert and A. Kemp and A. Nikroo and P. Antici and E. Brambrink and A. Blazevic and E. M. Campbell and J. C. Fernández and J.C. Gauthier and M. Geissel and M. Hegelich and H. Pépin and H. Popescu and N. Renard-LeGalloudec and M. Roth and J. Schreiber and R. Stephens and T. E. Cowan",
Journal = "{Phys. Rev. Lett.}",
Year = 2005,
Volume = 94,
Pages =  045004}

@article{fst07,
author = {T. Fukuhara and S. Sugawa and Y. Takahashi},
title = {Bose-Einstein condensation of an ytterbium isotope},
year = {2007},
journal = {Phys. Rev. A},
volume = {76},
eid = {051604(R)},
pages = {051604(R)}
}

@article{frn19,
  author = {Franssen, J. G. H. and de Raadt, T. C. H. and van Ninhuijs, M. A. W. and Luiten, O. J.},
  journal = {Phys. Rev. Accel. Beams},
  volume = {22},
  issue = {2},
  pages = {023401},
  numpages = {11},
  year = {2019},
  month = {Feb},
  publisher = {American Physical Society},
}

@article{fzr06,
  title = {Observation of Collective Modes of Ultracold Plasmas},
  author = {Fletcher, R. S. and Zhang, X. L. and Rolston, S. L.},
  journal = {Phys. Rev. Lett.},
  volume = {96},
  issue = {10},
  pages = {105003},
  numpages = {4},
  year = {2006},
  month = {Mar},
  publisher = {American Physical Society},
  doi = {10.1103/PhysRevLett.96.105003},
}


@article{fzr07,
  title = {Using Three-Body Recombination to Extract Electron Temperatures of Ultracold Plasmas},
  author = {Fletcher, R. S. and Zhang, X. L. and Rolston, S. L.},
  journal = {Phys. Rev. Lett.},
  volume = {99},
  issue = {14},
  pages = {145001},
  numpages = {4},
  year = {2007},
  month = {Oct},
  publisher = {American Physical Society},
  doi = {10.1103/PhysRevLett.99.145001},
}


@article{gls07,
Author = "P. Gupta and S. Laha and C. E. Simien and H. Gao and J. Castro and T. C. Killian and T.  Pohl",
Title = {Electron-Temperature Evolution in Expanding Ultracold Neutral Plasmas},
Journal = "{ Phys. Rev. Lett.}",
Year = 2007,
Volume = 99,
Pages =  75005}

@article{glw20,
  author = {Gorman, G. M. and Langin, T. K. and Warrens, M. K. and Vrinceanu, D. and Killian, T. C.},
  title = {Combined molecular-dynamics and quantum-trajectories simulation of laser-driven, collisional systems},
  journal = {Phys. Rev. A},
  volume = {101},
  issue = {1},
  pages = {012710},
  numpages = {13},
  year = {2020},
  month = {Jan},
  publisher = {American Physical Society},
  doi = {10.1103/PhysRevA.101.012710},
}

@article{gmo06, author = {T. Grismayer and P. Mora}, title = {Influence of a finite initial ion density gradient on plasma expansion into a vacuum}, publisher = {AIP}, year = {2006}, journal =
{Phys. Plasmas}, volume = {13}, number = {3}, eid = {032103}, pages =
{032103}, keywords = {plasma density; plasma transport processes; plasma
flow; plasma thermodynamics; numerical analysis} 
}

@article{gms03, 
Author = "D. O. Gericke and M. S. Murillo and D. Semkat and M. Bonitz and D. Kremp",  
Journal = "{J. Phys. A}", 
Year = 2003, 
Volume = 36, 
Pages =  6087
}

@article{gsb02, 
Author = "D. O. Gericke and M. Schlanges and Th. Bornath",  
Journal = "{Phys. Rev. E}", 
Year = 2002, 
Volume = 65, 
Pages =  036406
}

@article{gwb21,
Author = "G. M. Gorman and M. K. Warrens and S. J. Bradshaw and T. C. Killian",
Title = {Magnetic Confinement of an Ultracold Neutral Plasma},
Journal = "{Phys. Rev. Lett.}",
Year = 2021,
Volume = 126,
Pages =  085002 
}

@article{hfd97,
Author = "S. Hamaguchi and R. T. Farouki and D. H. E. Dubin ",
Journal = "{Phys. Rev. E}",
Year = 1997,
Volume = 56,
Number = 4,
Pages = 4671}

@article{hss17,
  author = {Haenel, R. and Schulz-Weiling, M. and Sous, J. and Sadeghi, H. and Aghigh, M. and Melo, L. and Keller, J. S. and Grant, E. R.},
  journal = {Phys. Rev. A},
  volume = {96},
  issue = {2},
  pages = {023613},
  numpages = {13},
  year = {2017},
  month = {Aug},
  publisher = {American Physical Society}
}

@article{ich82,
Author = "S. Ichimaru",
Journal = "{Rev. of Mod. Phys.}",
Year = 1982,
Volume = 54,
Number = 4,
Pages =  1017}

@article{ICFCollab22,
Author = "H. Abu-Schwareb and \textit{et al.}",
Journal = "{Phys. Rev. Lett}",
Year = 2022,
Volume = 129,
Number = 075001}

@article{itb70,
Author = "H. Ikezi and R. J. Taylor and D. R. Baker",
Journal = "{Phys. Rev. Lett.}",
Year = 1970,
Volume = 25,
Number = 1,
Pages =  11 
}

@article{kde87, 
Author = "J. T. Karpen and C. R. DeVore", 
Journal = "{ApJ}", 
Year = 1987, 
Volume = 320, 
Pages = 904
}

@article{klk01,
Author = "T. C. Killian and M. J. Lim and S. Kulin and R. Dumke and S. D. Bergeson and S. L. Rolston",
Title = {Formation of Rydberg Atoms in an Expanding Ultracold Neutral Plasma},
Journal = "{ Phys. Rev. Lett.}",
Year = 2001,
Volume = 86,
Number = 17,
Pages =  3759}

@article{kkb99,
Author = "T. C. Killian and S. Kulin and S. D. Bergeson and L. A. Orozco and C. Orzel and S. L. Rolston",
Title = {Creation of an Ultracold Neutral Plasma},
Journal = "{Phys. Rev. Lett.}",
Year = 1999,
Volume = 83,
Number = 23,
Pages =  4776}

@article{kmo12,
author = {Killian, T. C. and McQuillen, P. and O'Neil, T. M. and Castro, J.},
title = {Creating and studying ion acoustic waves in ultracold neutral plasmas},
doi = {https://doi.org/10.1063/1.3694654},
issn = {0031-9007},
journal = {Phys. Plasmas},
month = {march},
pages = {055701},
volume = {19},
year = {2012}
}

@article{kgs21,
author = {Kroker, T. and Großmann, M. and Sengstock, K. and Drescher, M. and Wessels-Staarmann, P. and Simonet, J. },
title = {Ultrafast electron cooling in an expanding ultracold plasma},
doi = {https://doi.org/10.1038/s41467-020-20815-8},
issn = {},
journal = {Nature Comm.},
month = {January},
pages = {596},
volume = {12},
year = {2021}
}

@article{kkb00,
Author = "S. Kulin and T. C. Killian and S. D. Bergeson and S. L. Rolston",
Title = {Plasma Oscillations and Expansion of an Ultracold Neutral Plasma},
Journal = "{Phys. Rev. Lett.}",
Year = 2000,
Volume = 85,
Number = 2,
Pages =  318
}



@article{kon02,
Author = "S. G. Kuzmin and T. M. O'Neil",
Journal = "{Phys. Plasmas}",
Year = 2002,
Volume = 9,
Number = 9,
Pages =  3743}

@article{kpp07,
  AUTHOR =       {T. C. Killian and T. Pattard and T. Pohl and J. M. Rost},
  Title = {Ultracold Neutral Plasmas},
 Journal = "{Phys. Rep.}",
Year = 2007,
Volume = 449,
Pages =  77}

@article{lbh15,
  author={Lyon, M and Bergeson, Scott D and Hart, G and Murillo, MS},
  title = {Strongly-coupled plasmas formed from laser-heated solids},
  journal={Sci. Rep.},
  volume={5},
  number={1},
  pages={1},
  year={2015},
  publisher={Nature Publishing Group}
}

@article{lbm13,
author = {Lyon, M. and Bergeson, S. D. and Murillo, M. S.},
doi = {10.1103/PhysRevE.87.033101},
issn = {1539-3755},
journal = {Phys. Rev. E},
month = {mar},
number = {3},
pages = {033101},
publisher = {American Physical Society},
volume = {87},
year = {2013}
}

@article{lcg06,
Author = "S. Laha and Y. C. Chen and P. Gupta and C. E. Simien and Y. N. Martinez and P. G. Mickelson and S. B. Nagel and T. C. Killian",
Title = {Kinetic energy oscillations in annular regions of ultracold neutral plasmas},
Journal = "{Eur. Phys. J. D}",
Year = 2006,
Volume = 40,
Pages =  51}

@ARTICLE{lgk19,
  AUTHOR =       {Thomas K. Langin and Grant M. Gorman and Thomas C. Killian},
  title = {Laser cooling of ions in a neutral plasma},
  JOURNAL =      {Science},
  YEAR =         {2019},
  volume =       {363},
  number =       {6422},
  pages =        {61-64},
}

@article{lgs07,
Author = "S. Laha and  P. Gupta and C. E. Simien and  H. Gao and J. Castro  and T. C. Killian",
Title = {Experimental Realization of an Exact Solution to the Vlasov Equations for an Expanding Plasma},
Journal = "{Phys. Rev. Lett.}",
Year = 2007,
Volume = 99,
Pages =  155001
}

@BOOK{ll87,
  AUTHOR =       {L. D. Landau and E. M. Lifshits},
  TITLE =        {Fluid Mechanics},
  PUBLISHER =    {Pergamon Press},
  YEAR =         {1987},
  address =      {Elmsford, New York},
}

@article{lin95,
Author = "J. Lindl",
Journal = "{Phys. Plasmas}",
Year = 1995,
Volume = 2,
Pages =  3933
}

@article{lro17,
author = {Lyon, M. and Rolston, S. L.},
journal = {Rep. Prog. Phys.},
pages = {017001},
volume = {80},
year = {2017}
}

@article{lsm16,
author = {Langin, T. K. and Strickler, T. and Maksimovic, N. and McQuillen, P. and Pohl, T. and Vrinceanu, D. and Killian, T. C.},
title = {Demonstrating universal scaling for dynamics of Yukawa one-component plasmas after an interaction quench},
journal = {Phys. Rev. E},
keywords = {Yukawa,ultracold neutral plasmas},
mendeley-tags = {Yukawa,ultracold neutral plasmas},
pages = {023201},
volume = {93},
year = {2016}
}

@article{mcb15,
author = {McQuillen, P. and Castro, J. and Bradshaw, S. J. and Killian, T. C.},
title = {Emergence of kinetic behavior in streaming ultracold neutral
plasmas},
issn = {1070-664X},
journal = {Phys. Plasmas},
keywords = {streaming plasmas,ultracold plasmas},
mendeley-tags = {streaming plasmas,ultracold plasmas},
pages = {043514},
volume = {22},
year = {2015}
}

@article{mck02, 
Author = "S. Mazevet and L. A. Collins and J. D. Kress",
Journal = "{Phys. Rev. Lett.}", 
Year = 2002, 
Volume = 88, 
Number = 5, 
Pages =  55001
}

@article{mcs13,
author = {McQuillen, P. and Castro, J. and Strickler, T. and Bradshaw, S. J. and Killian, T. C.},
title = {Ion holes in the hydrodynamic regime in ultracold neutral plasmas},
journal = {Phys. Plasmas},
month = {apr},
number = {4},
pages = {043516},
publisher = {AIP Publishing LLC},
volume = {20},
year = {2013}
}


@Article{mgf00,
  author = {A. Maksimchuk  and S. Gu  and K. Flippo and D. Umstadter and V. Yu. Bychenkov},
  journal = {Phys. Rev. Lett.},
  volume = {84},
  number = {18},
  pages = {4108},
  year = {2000},
  publisher = {American Physical Society}
}

@article{mke69,
Author = "P. Mansbach and J. Keck",
Journal = "{Phys. Rev.}",
Volume = 181,
Pages = 275,
Year = 1969}

@article{mch41,
Author = "D. McHenry",
Journal = "{J. Appl. Mech.}",
Volume = 8,
Year = 1941}

@article{mkn05, 
author = {M. Murakami and Y.-G. Kang and K. Nishihara and S.
Fujioka and H. Nishimura}, 
publisher = {AIP}, 
year ={2005}, 
journal = {Phys. Plasmas}, 
volume = {12}, number = {6}, 
eid = {062706}, 
pages = {062706} 
}

@article{mor05physplasmas, 
author = {P. Mora}, 
publisher = {AIP}, 
year = {2005}, 
journal = {Phys. Plasmas}, 
volume = {12}, 
number = {11}, 
eid = {112102}, 
pages = {112102}, 
keywords = {plasma simulation; fractals; plasma transport processes; plasma boundary layers}
}

@article{mro96,
  title = {Stopping power in nonideal and strongly coupled plasmas},
  author = {Morawetz, K. and R\"opke, G.},
  journal = {Phys. Rev. E},
  volume = {54},
  issue = {4},
  pages = {4134--4146},
  numpages = {0},
  year = {1996},
  month = {Oct},
  publisher = {American Physical Society},
  doi = {10.1103/PhysRevE.54.4134},
}


@article{mrk08, author = {J. P. Morrison and C. J. Rennick and J. S. Keller and E. R. Grant}, title = {Evolution from a Molecular Rydberg Gas to an Ultracold Plasma in a Seeded Supersonic Expansion of NO}, publisher = {APS}, year = {2008}, journal = {Phys. Rev. Lett.},
volume = {101}, number = {20}, eid = {205005}, numpages = {4}, pages =
{205005} 
}

@article{msc16,
author = {A. A. Mamun and R. Schlickeiser},
journal = {Phys. Plasmas},
pages = {034502},
volume = {23},
year = {2016}
}

@article{msk16,
author = {McClelland, J. J. and Steele, A. V. and Knuffman, B. and Twedt, K. A. and Schwarzkopf, A. and Wilson, T. M.},
issn = {19319401},
journal = {Appl. Phys. Rev.},
number = {1},
pages = {011302},
volume = {3},
year = {2016}
}

@article{msl15,
author = {McQuillen, P and Strickler, T. and Langin, T and Killian, T. C.},
title = {Experimental Measurement of Self-Diffusion in a Strongly Coupled Plasma},
journal = {Phys. Plasmas},
pages = {033513},
volume = {22},
year = {2015}
}

@article{mss11,
author = {McCulloch, A. J. and Sheludko, D. V. and Saliba, S. D. and Bell, S. C. and Junker, M. and Nugent, K. A. and Scholten, R. E.},
issn = {1745-2473},
journal = {Nature Physics},
keywords = {electron images,ultracold plasmas},
mendeley-tags = {electron images,ultracold plasmas},
month = {jul},
number = {7},
pages = {1},
publisher = {Nature Publishing Group},
volume = {7},
year = {2011}
}

@BOOK{mvs99,
  AUTHOR =       {H. J. Metcalf and P. van der Straten},
  TITLE =        {Laser Cooling and Trapping},
  PUBLISHER =    {Springer-Verlag},
  YEAR =         {1999},
  address =      {New York, New York},
}

@article{mgr15,
  title = {Dynamics of colliding ultracold plasmas},
  author = {Morrison, J. P. and Grant, E. R.},
  journal = {Phys. Rev. A},
  volume = {91},
  issue = {2},
  pages = {023423},
  numpages = {5},
  year = {2015},
  month = {Feb},
  publisher = {American Physical Society},
  doi = {10.1103/PhysRevA.91.023423},
}


@article{mur01,
Author = "M. S. Murillo",
Title = {Using Fermi Statistics to Create Strongly Coupled Ion Plasmas in Atom Traps},
Journal = "{Phys. Rev. Lett.}",
Year = 2001,
Volume = 87,
Number = 11,
Pages =  115003}

@article{mur04, author={M. S. Murillo},
         journal = {Phys. Plasmas}, volume=11, year=2004, pages=2964}


@article{mur06PRL,
Author = "M. S. Murillo",
Journal = "{Phys. Rev. Lett.}",
Year = 2006,
Volume = 96,
Pages =  165001}

@article{mur07PoP,
author = {Murillo, M. S.},
title = {Ultrafast dynamics of neutral, ultracold plasmas},
doi = {10.1063/1.2436853},
issn = {1070664X},
journal = {Phys. Plasmas},
keywords = {equilibration,molecular dynamics,plasma oscillations,plasma temperature,plasma transport processes,ultracold plasmas},
language = {en},
mendeley-tags = {equilibration,molecular dynamics,ultracold plasmas},
month = {mar},
number = {5},
pages = {055702},
volume = {14},
year = {2007}
}

@article{mzo14,
Author = "A. A. Mamun and M. S. Zobaer",
Journal = "{Phys. Plasmas}",
Year = 2014,
Volume = 21,
Number = 10,
Pages =  022101 
}

@article{nba99,
Author = "Y. Nakamura and H. Bailung",
Journal = "{Phys. Rev. Lett.}",
Year = 1999,
Volume = 83,
Number = 8,
Pages =  1602 
}

@article{nsl03,
Author = "S. B. Nagel and C. E. Simien and S. Laha and P. Gupta and V. S. Ashoka and T. C. Killian",
Title = {Magnetic trapping of metastable $^{3}\uppercase{\textnormal{P}}_{2}$ atomic strontium},
Journal = "{ Phys. Rev. A}",
Year = 2003,
Volume = 67,
eid =  {011401(R)},
Pages =  {011401(R)}}


@article{pdf13,
author = {Perrone, D. and Dendy, R. O. and Furno, I. and Sanchez, R. and Zimbardo, G. and Bovet, A. and Fasoli, A. and Gustafson, K. and Perri, S. and Ricci, P. and Valentini, F.},
doi = {10.1007/s11214-013-9966-9},
issn = {0038-6308},
journal = {Space Sci. Rev.},
keywords = {nonlocal transport,plasmas,space plasmas},
mendeley-tags = {nonlocal transport,plasmas,space plasmas},
month = {mar},
pages = {233},
volume = {178},
year = {2013}
}

@article{pmo94,
Author="M. D. Perry and G. Mourou",
Journal = "{Science}",
Year = 1994,
Volume = 264,
Pages = 917}

@article{ppr04PRA,
Author = "T. Pohl and T. Pattard  and J. M. Rost",
Title = {Kinetic modeling and molecular dynamics simulation of ultracold neutral plasmas including ionic correlations},
Journal = "{ Phys. Rev. A}",
Year = 2004,
Volume = 70,
Number = 3,
Pages =  033416
}

@article{ppr04PRL,
  title = {Coulomb Crystallization in Expanding Laser-Cooled Neutral Plasmas},
  author = {Pohl, T. and Pattard, T. and Rost, J. M.},
  journal = {Phys. Rev. Lett.},
  volume = {92},
  issue = {15},
  pages = {155003},
  numpages = {4},
  year = {2004},
  month = {Apr},
  publisher = {American Physical Society},
  doi = {10.1103/PhysRevLett.92.155003},
}


@article{ppr05PRL,
Author = "T. Pohl and T. Pattard  and J. M. Rost",
Journal = "{ Phys. Rev. Lett.}",
Year = 2005,
Volume = 94,
Pages =  205003
}

@article{pvs08,
    Author = {T. Pohl and D. Vrinceanu and H. R. Sadeghpour},
    Journal = {Phys. Rev. Lett.},
    Pages = {223201},
    Volume = {100},
    Year = {2008}
}

@article{rfb05,
Author = "L. Romagnani and J. Fuchs and M. Borghesi and P. Antici and P. Audebert and F. Ceccherini and T. Cowan and T. Grismayer and S. Kar and A. Macchi and P. Mora and G. Pretzler and A. Schiavi and T. Toncian and O. Willi",
Journal = "{Phys. Rev. Lett.}",
Year = 2005,
Volume = 95,
Pages =  195001
}

@article{rbb08,
Author = "L. Romagnani and S.V. Bulanov and M. Borghesi adn P. Audebert and J.C. Bauthier and K. Lowenbruck and A.J. Mackinnon adn P. Patel adn G. Pretzler and T. Toncian and O. Willi",
Journal = "{Phys. Rev. Lett.}",
Year = 2008,
Volume = 101,
Pages =  025004
}

@article{rha02,
Author = "F. Robicheaux and J. D. Hanson",
Title = {Simulation of the Expansion of an Ultracold Neutral Plasma},
Journal = "{Phys. Rev. Lett.}",
Year = 2002,
Volume = 88,
Number = 5,
Pages =  55002
}

@article{rha03, 
Author = "F. Robicheaux and J. D. Hanson",  
Title = {Simulated expansion of an ultra-cold, neutral plasma},
Journal = "{ Phys. Plasmas}", 
Year = 2003, 
Volume = 10, 
Number = 6, 
Pages =  2217}

@article{ssc85,
doi = {10.1088/0741-3335/27/7/002},
year = {1985},
month = {jul},
publisher = {},
volume = {27},
number = {7},
pages = {717},
author = {C Sack and  H Schamel},
title = {Evolution of a plasma expanding into vacuum},
journal = {Plasma Physics and Controlled Fusion},
}


@ARTICLE{sno67,
  author={Snow, W.},
  journal={IEEE Transactions on Audio and Electroacoustics}, 
  title={Survey of acoustic characteristics of bullet shock waves}, 
  year={1967},
  volume={15},
  number={4},
  pages={161-176},
  doi={10.1109/TAU.1967.1161921}}

@article{shh07,
author = {D. R. Symes and M. Hohenberger and A. Henig and T. Ditmire},
publisher = {APS},
year = {2007},
journal = {Phys. Rev. Lett.},
volume = {98},
number = {12},
eid = {123401},
pages = {123401},
}

@Article{rkt09,
  author = {Reijnders, M. P. and van Kruisbergen, P. A. and Taban, G.  and van der Geer, S. B. and Mutsaers, P. H. A. and Vredenbregt, E. J. D. and Luiten, O. J.},
  journal = {Phys. Rev. Lett.},
  volume = {102},
  number = {3},
  pages = {034802},
  numpages = {4},
  year = {2009},
  doi = {10.1103/PhysRevLett.102.034802},
  publisher = {American Physical Society}
}

@article{rpc87,
Author = "E. L. Raab and M. Prentiss and A. Cable and S. Chu and D. E. Pritchard",
Journal = "{ Phys. Rev. Lett.}",
Year = 1987,
Volume = 59,
Number = 23,
Pages =  2631}

@article{sbs22,
Author = " R. T. Sprenkle and S. D. Bergeson and L. G. Silvestri and M. S. Murillo",
Title = {Ultracold neutral plasma expansion in a strong uniform magnetic field},
Journal = "{Phys. Rev. E}",
Year = 2022,
Volume = {105},
Pages = {045201}
}

@article{sbl22,
Author = " R. B. Scott  and S. J. Bradshaw and M. G. Linton",
title = {The Dynamic Evolution of Solar Wind Streams Following Interchange Reconnection},
Journal = "{Astrophys. J.}",
Year = 2022,
Volume = {933},
Pages = {72}
}

@article{scg04,
Author = " C. E. Simien and Y. C. Chen and P. Gupta and S. Laha and Y. N. Martinez and P. G. Mickelson and S. B. Nagel and T. C. Killian",
Journal = "{Phys. Rev. Lett.}",
Year = 2004,
Volume = {92},
number={14},
Pages = {143001}
}

@article{sgr12,
author = {Sadeghi, H. and Grant, E. R.},
doi = {10.1103/PhysRevA.86.052701},
issn = {1050-2947},
journal = {Phys. Rev. A},
keywords = {molecular beam plasmas,ultracold plasmas},
mendeley-tags = {molecular beam plasmas,ultracold plasmas},
month = {nov},
number = {5},
pages = {052701},
publisher = {American Physical Society},
volume = {86},
year = {2012}
}

@article{sma13,
    author = {Shahmansouri, M. and Mamun, A. A.},
    title = "{Oblique ion acoustic shock waves in a magnetized plasma}",
    journal = {Physics of Plasmas},
    volume = {20},
    number = {8},
    pages = {082122},
    year = {2013},
    month = {08},
    issn = {1070-664X},
    doi = {10.1063/1.4818492},
    eprint = {https://pubs.aip.org/aip/pop/article-pdf/doi/10.1063/1.4818492/16052780/082122\_1\_online.pdf},
}


@Article{skh00,
  author = {Snavely, R. A. and Key, M. H. and Hatchett, S. P. and Cowan, T. E. and Roth, M.  and Phillips, T. W. and Stoyer, M. A. and Henry, E. A. and Sangster, T. C. and Singh, M. S. and Wilks, S. C. and MacKinnon, A.  and Offenberger, A.  and Pennington, D. M. and Yasuike, K.  and Langdon, A. B. and Lasinski, B. F. and Johnson, J.  and Perry, M. D. and Campbell, E. M.},
  journal = {Phys. Rev. Lett.},
  volume = {85},
  number = {14},
  pages = {2945},
  numpages = {3},
  year = {2000},
  doi = {10.1103/PhysRevLett.85.2945},
  publisher = {American Physical Society}
}

@article{sdm19,
  title = {Ion friction at small values of the Coulomb logarithm},
  author = {Sprenkle, Tucker and Dodson, Adam and McKnight, Emma and Spencer, Ross and Bergeson, Scott and Diaw, Abdourahmane and Murillo, Michael S.},
  journal = {Phys. Rev. E},
  volume = {99},
  issue = {5},
  pages = {053206},
  numpages = {8},
  year = {2019},
  month = {May},
  publisher = {American Physical Society},
  doi = {10.1103/PhysRevE.99.053206},
}


@article{slm16,
author = {Strickler, T. S. and Langin, T. K. and McQuillen, P. and Daligault, J. and Killian, T. C.},
journal = {Phys. Rev. X},
keywords = {diffusion,ultracold neutral plasmas},
mendeley-tags = {diffusion,ultracold neutral plasmas},
pages = {021021},
volume = {6},
year = {2016}
}

@article{smd11,
Author = "P.K. Shukla and W. M. Moslem and S. S. Duha and A. A. Mamun",
Title = {Time evolution of cylindrical and spherical shock waves in an ultracold neutral plasma with non-Maxwellian electrons},
Journal = "{Europhys. Lett}",
Year = 2011,
Volume = 96,
Number = 6,
Pages =  65002 
}

@article{ssm21,
author = {Sun, Yang-Yi and Shen, Mitchell M. and Tsai, Yu-Lin and Lin, Chi-Yen and Chou, Min-Yang and Yu, Tao and Lin, Kai and Huang, Qian and Wang, Jin and Qiu, Lihui and Chen, Chieh-Hung and Liu, Jann-Yenq},
title = {Wave Steepening in Ionospheric Total Electron Density due to the 21 August 2017 Total Solar Eclipse},
journal = {Journal of Geophysical Research: Space Physics},
volume = {126},
number = {3},
pages = {e2020JA028931},
keywords = {GNSS, nonlinear wave, solar eclipse, total electron content, wave break, wave steepening},
doi = {https://doi.org/10.1029/2020JA028931},
year = {2021}
}

@article{tre09,
author = {R. A. Treumann},
title = {Fundamentals of Collisionless Shocks for Astrophysical Application, 1. Non-relativistic Shocks},
journal = {The Astronomy and Astrophysics Review},
volume = {17},
pages = {409},
doi = {https://10.1007/s00159-009-0024-2},
year = {2009}
}

@article{uwc87,
  title = {Observation of steepening in electron plasma waves driven by stimulated Raman backscattering},
  author = {Umstadter, D. and Williams, R. and Clayton, C. and Joshi, C.},
  journal = {Phys. Rev. Lett.},
  volume = {59},
  issue = {3},
  pages = {292--295},
  numpages = {0},
  year = {1987},
  month = {Jul},
  publisher = {American Physical Society},
  doi = {10.1103/PhysRevLett.59.292},
}


@article{vmr20,
Author = "M. A. Viray and S. A. Miller and G. Raithel",
Title = {Coulomb expansion of a cold non-neutral rubidium plasma},
Journal = "{Phys. Rev. A}",
Year = 2020,
Volume = 102,
Pages = 033303
}

@article{vbk20,
author = {Vikhrov,E. V.  and Bronin, S. Ya.  and Klayrfeld, A. B.  and Zelener,B. B.  and Zelener,B. V. },
journal = {Phys. Plasmas},
volume = {27},
number = {12},
pages = {120702},
year = {2020},
}

@article{vbz21,
Author = "E.V. Vikhrov and S. YA. Bronin and B. B. Zelener and B. V. Zelener",
Title = {Ion wave formation during ultracold plasma expansion},
Journal = "{Phys. Rev. E}",
Year = 2021,
Volume = 104,
Pages =  015212 
}

@article{wbc08,
author = {West, M. J. and Bradshaw, S. J. and Cargill, P. J.},
doi = {10.1007/s11207-008-9243-3},
isbn = {1120700892},
issn = {0038-0938},
journal = {Sol. Phys.},
keywords = {nonlocal transport,plasma,solar atmosphere},
mendeley-tags = {nonlocal transport,plasma,solar atmosphere},
month = {aug},
number = {1},
pages = {89},
volume = {252},
year = {2008}
}

@article{wgb21,
Author = "M. K. Warrens and G. M. Gorman and S. J. Bradshaw and T. C. Killian",
Title = {Expansion of ultracold neutral plasmas with exponentially decaying density distributions},
Journal = "{Phys. Plasmas}",
Year = 2021,
Volume = 28,
Pages =  022110 
}

@book{whi99,
author = {G. B. Whitham},
address = {New York :},
booktitle = {Linear and nonlinear waves},
edition = {1},
isbn = {0-471-35942-4},
lccn = {99028957},
publisher = {Wiley-Interscience,},
title = {Introduction and General Outline},
year = {1999},
}

@article{xlh03,
author = {Xinye Xu and Thomas H. Loftus and John L. Hall and Alan Gallagher and Jun Ye},
journal = {J. Opt. Soc. Am. B},
number = {5},
pages = {968--976},
publisher = {Optica Publishing Group},
title = {Cooling and trapping of atomic strontium},
volume = {20},
month = {May},
year = {2003},
doi = {10.1364/JOSAB.20.000968}
}

@Article{zbv22,
  author = {B. B. Zelener and S. YA. Bronin and E. V. Vilshanskaya and E. V. Vikhrov and K. P. Galstyan and N. V. Morozov and S. A. Saakyan and V. A. Sautenkof and B. V. Zelener},
  journal = {Quantum Electronics},
  volume = {52},
  number = {6},
  pages = {523},
  year = {2022},
}

@article{bvz23,
  title = {Ultracold plasma expansion in quadrupole magnetic field},
  author = {Bronin, S. Ya. and Vikhrov, E. V. and Zelener, B. B. and Zelener, B. V.},
  journal = {Phys. Rev. E},
  volume = {108},
  issue = {4},
  pages = {045209},
  numpages = {5},
  year = {2023},
  month = {Oct},
  publisher = {American Physical Society},
  doi = {10.1103/PhysRevE.108.045209},
}


@article{zfr08instability,
  title = {Observation of an Ultracold Plasma Instability},
  author = {Zhang, X. L. and Fletcher, R. S. and Rolston, S. L.},
  journal = {Phys. Rev. Lett.},
  volume = {101},
  issue = {19},
  pages = {195002},
  numpages = {4},
  year = {2008},
  month = {Nov},
  publisher = {American Physical Society},
  doi = {10.1103/PhysRevLett.101.195002},
}



@article{zfr08Bfield,
  title = {Ultracold Plasma Expansion in a Magnetic Field},
  author = {Zhang, X. L. and Fletcher, R. S. and Rolston, S. L. and Guzdar, P. N. and Swisdak, M.},
  journal = {Phys. Rev. Lett.},
  volume = {100},
  issue = {23},
  pages = {235002},
  numpages = {4},
  year = {2008},
  month = {Jun},
  publisher = {American Physical Society},
  doi = {10.1103/PhysRevLett.100.235002},
}

@INPROCEEDINGS{Szypko2023Modeling,
	author = {Szypko, Gregory and Gorman, Grant and Bradshaw, Stephen},
	journal = {Bulletin of the AAS},
	number = {7},
	year = {2023},
	month = {sep 18},
	note = {https://baas.aas.org/pub/2023n7i107p04},
	publisher = {},
	title = {Modeling {Fluid} {Evolution} of {Interchange} {Reconnection} {Aftermath}},
	volume = {55},
    note = {A POSTER Presentation}
}

@article{Centrella1984,
	author = {Centrella, Joan and Wilson, James W.},
	journal = {The Astrophysical Journal Suplement Series},
	volume = {54},
	year = {1984},
	month = {Feb},
    pages = {229-249},
	publisher = {American Astronomical Society},
	title = {Planar Numerical Cosmology. II. The Difference Equations and Numerical Tests},
}



\begin{thebibliography}{50}%
\makeatletter
\providecommand \@ifxundefined [1]{%
 \@ifx{#1\undefined}
}%
\providecommand \@ifnum [1]{%
 \ifnum #1\expandafter \@firstoftwo
 \else \expandafter \@secondoftwo
 \fi
}%
\providecommand \@ifx [1]{%
 \ifx #1\expandafter \@firstoftwo
 \else \expandafter \@secondoftwo
 \fi
}%
\providecommand \natexlab [1]{#1}%
\providecommand \enquote  [1]{``#1''}%
\providecommand \bibnamefont  [1]{#1}%
\providecommand \bibfnamefont [1]{#1}%
\providecommand \citenamefont [1]{#1}%
\providecommand \href@noop [0]{\@secondoftwo}%
\providecommand \href [0]{\begingroup \@sanitize@url \@href}%
\providecommand \@href[1]{\@@startlink{#1}\@@href}%
\providecommand \@@href[1]{\endgroup#1\@@endlink}%
\providecommand \@sanitize@url [0]{\catcode `\\12\catcode `\$12\catcode
  `\&12\catcode `\#12\catcode `\^12\catcode `\_12\catcode `\%12\relax}%
\providecommand \@@startlink[1]{}%
\providecommand \@@endlink[0]{}%
\providecommand \url  [0]{\begingroup\@sanitize@url \@url }%
\providecommand \@url [1]{\endgroup\@href {#1}{\urlprefix }}%
\providecommand \urlprefix  [0]{URL }%
\providecommand \Eprint [0]{\href }%
\providecommand \doibase [0]{http://dx.doi.org/}%
\providecommand \selectlanguage [0]{\@gobble}%
\providecommand \bibinfo  [0]{\@secondoftwo}%
\providecommand \bibfield  [0]{\@secondoftwo}%
\providecommand \translation [1]{[#1]}%
\providecommand \BibitemOpen [0]{}%
\providecommand \bibitemStop [0]{}%
\providecommand \bibitemNoStop [0]{.\EOS\space}%
\providecommand \EOS [0]{\spacefactor3000\relax}%
\providecommand \BibitemShut  [1]{\csname bibitem#1\endcsname}%
\let\auto@bib@innerbib\@empty
\bibitem [{\citenamefont {Landau}\ and\ \citenamefont {Lifshits}(1987)}]{ll87}%
  \BibitemOpen
  \bibfield  {author} {\bibinfo {author} {\bibfnamefont {L.~D.}\ \bibnamefont
  {Landau}}\ and\ \bibinfo {author} {\bibfnamefont {E.~M.}\ \bibnamefont
  {Lifshits}},\ }\href@noop {} {\emph {\bibinfo {title} {Fluid Mechanics}}}\
  (\bibinfo  {publisher} {Pergamon Press},\ \bibinfo {address} {Elmsford, New
  York},\ \bibinfo {year} {1987})\BibitemShut {NoStop}%
\bibitem [{\citenamefont {Balogh}\ and\ \citenamefont
  {Treumann}(2013)}]{btr13}%
  \BibitemOpen
  \bibfield  {author} {\bibinfo {author} {\bibfnamefont {A.}~\bibnamefont
  {Balogh}}\ and\ \bibinfo {author} {\bibfnamefont {R.~A.}\ \bibnamefont
  {Treumann}},\ }\href@noop {} {\emph {\bibinfo {title} {Physics of
  Collisionless Shocks}}}\ (\bibinfo  {publisher} {Springer},\ \bibinfo
  {address} {New York, New York},\ \bibinfo {year} {2013})\BibitemShut
  {NoStop}%
\bibitem [{\citenamefont {Whitham}(1999)}]{whi99}%
  \BibitemOpen
  \bibfield  {author} {\bibinfo {author} {\bibfnamefont {G.~B.}\ \bibnamefont
  {Whitham}},\ }\href@noop {} {\emph {\bibinfo {title} {Linear and nonlinear
  waves}}},\ \bibinfo {edition} {1st}\ ed.\ (\bibinfo  {publisher}
  {Wiley-Interscience,},\ \bibinfo {address} {New York :},\ \bibinfo {year}
  {1999})\BibitemShut {NoStop}%
\bibitem [{\citenamefont {Campanella}\ \emph {et~al.}(2019)\citenamefont
  {Campanella}, \citenamefont {Legnaioli}, \citenamefont {Pagnotta},
  \citenamefont {Poggialini},\ and\ \citenamefont {Palleschi}}]{clp19}%
  \BibitemOpen
  \bibfield  {author} {\bibinfo {author} {\bibfnamefont {B.}~\bibnamefont
  {Campanella}}, \bibinfo {author} {\bibfnamefont {S.}~\bibnamefont
  {Legnaioli}}, \bibinfo {author} {\bibfnamefont {S.}~\bibnamefont {Pagnotta}},
  \bibinfo {author} {\bibfnamefont {F.}~\bibnamefont {Poggialini}}, \ and\
  \bibinfo {author} {\bibfnamefont {V.}~\bibnamefont {Palleschi}},\ }\href
  {https://www.mdpi.com/2218-2004/7/2/57} {\bibfield  {journal} {\bibinfo
  {journal} {Atoms}\ }\textbf {\bibinfo {volume} {7}} (\bibinfo {year}
  {2019})}\BibitemShut {NoStop}%
\bibitem [{\citenamefont {Treumann}(2009)}]{tre09}%
  \BibitemOpen
  \bibfield  {author} {\bibinfo {author} {\bibfnamefont {R.~A.}\ \bibnamefont
  {Treumann}},\ }\href {\doibase https://10.1007/s00159-009-0024-2} {\bibfield
  {journal} {\bibinfo  {journal} {The Astronomy and Astrophysics Review}\
  }\textbf {\bibinfo {volume} {17}},\ \bibinfo {pages} {409} (\bibinfo {year}
  {2009})}\BibitemShut {NoStop}%
\bibitem [{\citenamefont {Scott}\ \emph {et~al.}(2022)\citenamefont {Scott},
  \citenamefont {Bradshaw},\ and\ \citenamefont {Linton}}]{sbl22}%
  \BibitemOpen
  \bibfield  {author} {\bibinfo {author} {\bibfnamefont {R.~B.}\ \bibnamefont
  {Scott}}, \bibinfo {author} {\bibfnamefont {S.~J.}\ \bibnamefont {Bradshaw}},
  \ and\ \bibinfo {author} {\bibfnamefont {M.~G.}\ \bibnamefont {Linton}},\
  }\href@noop {} {\bibfield  {journal} {\bibinfo  {journal} {{Astrophys. J.}}\
  }\textbf {\bibinfo {volume} {933}},\ \bibinfo {pages} {72} (\bibinfo {year}
  {2022})}\BibitemShut {NoStop}%
\bibitem [{\citenamefont {Killian}\ \emph {et~al.}(1999)\citenamefont
  {Killian}, \citenamefont {Kulin}, \citenamefont {Bergeson}, \citenamefont
  {Orozco}, \citenamefont {Orzel},\ and\ \citenamefont {Rolston}}]{kkb99}%
  \BibitemOpen
  \bibfield  {author} {\bibinfo {author} {\bibfnamefont {T.~C.}\ \bibnamefont
  {Killian}}, \bibinfo {author} {\bibfnamefont {S.}~\bibnamefont {Kulin}},
  \bibinfo {author} {\bibfnamefont {S.~D.}\ \bibnamefont {Bergeson}}, \bibinfo
  {author} {\bibfnamefont {L.~A.}\ \bibnamefont {Orozco}}, \bibinfo {author}
  {\bibfnamefont {C.}~\bibnamefont {Orzel}}, \ and\ \bibinfo {author}
  {\bibfnamefont {S.~L.}\ \bibnamefont {Rolston}},\ }\href@noop {} {\bibfield
  {journal} {\bibinfo  {journal} {{Phys. Rev. Lett.}}\ }\textbf {\bibinfo
  {volume} {83}},\ \bibinfo {pages} {4776} (\bibinfo {year}
  {1999})}\BibitemShut {NoStop}%
\bibitem [{\citenamefont {Killian}\ \emph {et~al.}(2007)\citenamefont
  {Killian}, \citenamefont {Pattard}, \citenamefont {Pohl},\ and\ \citenamefont
  {Rost}}]{kpp07}%
  \BibitemOpen
  \bibfield  {author} {\bibinfo {author} {\bibfnamefont {T.~C.}\ \bibnamefont
  {Killian}}, \bibinfo {author} {\bibfnamefont {T.}~\bibnamefont {Pattard}},
  \bibinfo {author} {\bibfnamefont {T.}~\bibnamefont {Pohl}}, \ and\ \bibinfo
  {author} {\bibfnamefont {J.~M.}\ \bibnamefont {Rost}},\ }\href@noop {}
  {\bibfield  {journal} {\bibinfo  {journal} {{Phys. Rep.}}\ }\textbf {\bibinfo
  {volume} {449}},\ \bibinfo {pages} {77} (\bibinfo {year} {2007})}\BibitemShut
  {NoStop}%
\bibitem [{\citenamefont {Kroker}\ \emph {et~al.}(2021)\citenamefont {Kroker},
  \citenamefont {Großmann}, \citenamefont {Sengstock}, \citenamefont
  {Drescher}, \citenamefont {Wessels-Staarmann},\ and\ \citenamefont
  {Simonet}}]{kgs21}%
  \BibitemOpen
  \bibfield  {author} {\bibinfo {author} {\bibfnamefont {T.}~\bibnamefont
  {Kroker}}, \bibinfo {author} {\bibfnamefont {M.}~\bibnamefont {Großmann}},
  \bibinfo {author} {\bibfnamefont {K.}~\bibnamefont {Sengstock}}, \bibinfo
  {author} {\bibfnamefont {M.}~\bibnamefont {Drescher}}, \bibinfo {author}
  {\bibfnamefont {P.}~\bibnamefont {Wessels-Staarmann}}, \ and\ \bibinfo
  {author} {\bibfnamefont {J.}~\bibnamefont {Simonet}},\ }\href {\doibase
  https://doi.org/10.1038/s41467-020-20815-8} {\bibfield  {journal} {\bibinfo
  {journal} {Nature Comm.}\ }\textbf {\bibinfo {volume} {12}},\ \bibinfo
  {pages} {596} (\bibinfo {year} {2021})}\BibitemShut {NoStop}%
\bibitem [{\citenamefont {Morrison}\ \emph {et~al.}(2008)\citenamefont
  {Morrison}, \citenamefont {Rennick}, \citenamefont {Keller},\ and\
  \citenamefont {Grant}}]{mrk08}%
  \BibitemOpen
  \bibfield  {author} {\bibinfo {author} {\bibfnamefont {J.~P.}\ \bibnamefont
  {Morrison}}, \bibinfo {author} {\bibfnamefont {C.~J.}\ \bibnamefont
  {Rennick}}, \bibinfo {author} {\bibfnamefont {J.~S.}\ \bibnamefont {Keller}},
  \ and\ \bibinfo {author} {\bibfnamefont {E.~R.}\ \bibnamefont {Grant}},\
  }\href@noop {} {\bibfield  {journal} {\bibinfo  {journal} {Phys. Rev. Lett.}\
  }\textbf {\bibinfo {volume} {101}},\ \bibinfo {eid} {205005} (\bibinfo {year}
  {2008})}\BibitemShut {NoStop}%
\bibitem [{\citenamefont {Kulin}\ \emph {et~al.}(2000)\citenamefont {Kulin},
  \citenamefont {Killian}, \citenamefont {Bergeson},\ and\ \citenamefont
  {Rolston}}]{kkb00}%
  \BibitemOpen
  \bibfield  {author} {\bibinfo {author} {\bibfnamefont {S.}~\bibnamefont
  {Kulin}}, \bibinfo {author} {\bibfnamefont {T.~C.}\ \bibnamefont {Killian}},
  \bibinfo {author} {\bibfnamefont {S.~D.}\ \bibnamefont {Bergeson}}, \ and\
  \bibinfo {author} {\bibfnamefont {S.~L.}\ \bibnamefont {Rolston}},\
  }\href@noop {} {\bibfield  {journal} {\bibinfo  {journal} {{Phys. Rev.
  Lett.}}\ }\textbf {\bibinfo {volume} {85}},\ \bibinfo {pages} {318} (\bibinfo
  {year} {2000})}\BibitemShut {NoStop}%
\bibitem [{\citenamefont {Bergeson}\ and\ \citenamefont
  {Spencer}(2003)}]{bsp03}%
  \BibitemOpen
  \bibfield  {author} {\bibinfo {author} {\bibfnamefont {S.~D.}\ \bibnamefont
  {Bergeson}}\ and\ \bibinfo {author} {\bibfnamefont {R.~L.}\ \bibnamefont
  {Spencer}},\ }\href {\doibase 10.1103/PhysRevE.67.026414} {\bibfield
  {journal} {\bibinfo  {journal} {Phys. Rev. E}\ }\textbf {\bibinfo {volume}
  {67}},\ \bibinfo {pages} {026414} (\bibinfo {year} {2003})}\BibitemShut
  {NoStop}%
\bibitem [{\citenamefont {Castro}\ \emph {et~al.}(2010)\citenamefont {Castro},
  \citenamefont {McQuillen},\ and\ \citenamefont {Killian}}]{cmk10}%
  \BibitemOpen
  \bibfield  {author} {\bibinfo {author} {\bibfnamefont {J.}~\bibnamefont
  {Castro}}, \bibinfo {author} {\bibfnamefont {P.}~\bibnamefont {McQuillen}}, \
  and\ \bibinfo {author} {\bibfnamefont {T.~C.}\ \bibnamefont {Killian}},\
  }\href@noop {} {\bibfield  {journal} {\bibinfo  {journal} {{Phys. Rev.
  Lett.}}\ }\textbf {\bibinfo {volume} {105}},\ \bibinfo {pages} {065004}
  (\bibinfo {year} {2010})}\BibitemShut {NoStop}%
\bibitem [{\citenamefont {McQuillen}\ \emph {et~al.}(2013)\citenamefont
  {McQuillen}, \citenamefont {Castro}, \citenamefont {Strickler}, \citenamefont
  {Bradshaw},\ and\ \citenamefont {Killian}}]{mcs13}%
  \BibitemOpen
  \bibfield  {author} {\bibinfo {author} {\bibfnamefont {P.}~\bibnamefont
  {McQuillen}}, \bibinfo {author} {\bibfnamefont {J.}~\bibnamefont {Castro}},
  \bibinfo {author} {\bibfnamefont {T.}~\bibnamefont {Strickler}}, \bibinfo
  {author} {\bibfnamefont {S.~J.}\ \bibnamefont {Bradshaw}}, \ and\ \bibinfo
  {author} {\bibfnamefont {T.~C.}\ \bibnamefont {Killian}},\ }\href@noop {}
  {\bibfield  {journal} {\bibinfo  {journal} {Phys. Plasmas}\ }\textbf
  {\bibinfo {volume} {20}},\ \bibinfo {pages} {043516} (\bibinfo {year}
  {2013})}\BibitemShut {NoStop}%
\bibitem [{\citenamefont {McQuillen}\ \emph
  {et~al.}(2015{\natexlab{a}})\citenamefont {McQuillen}, \citenamefont
  {Castro}, \citenamefont {Bradshaw},\ and\ \citenamefont {Killian}}]{mcb15}%
  \BibitemOpen
  \bibfield  {author} {\bibinfo {author} {\bibfnamefont {P.}~\bibnamefont
  {McQuillen}}, \bibinfo {author} {\bibfnamefont {J.}~\bibnamefont {Castro}},
  \bibinfo {author} {\bibfnamefont {S.~J.}\ \bibnamefont {Bradshaw}}, \ and\
  \bibinfo {author} {\bibfnamefont {T.~C.}\ \bibnamefont {Killian}},\
  }\href@noop {} {\bibfield  {journal} {\bibinfo  {journal} {Phys. Plasmas}\
  }\textbf {\bibinfo {volume} {22}},\ \bibinfo {pages} {043514} (\bibinfo
  {year} {2015}{\natexlab{a}})}\BibitemShut {NoStop}%
\bibitem [{\citenamefont {Killian}\ \emph {et~al.}(2001)\citenamefont
  {Killian}, \citenamefont {Lim}, \citenamefont {Kulin}, \citenamefont {Dumke},
  \citenamefont {Bergeson},\ and\ \citenamefont {Rolston}}]{klk01}%
  \BibitemOpen
  \bibfield  {author} {\bibinfo {author} {\bibfnamefont {T.~C.}\ \bibnamefont
  {Killian}}, \bibinfo {author} {\bibfnamefont {M.~J.}\ \bibnamefont {Lim}},
  \bibinfo {author} {\bibfnamefont {S.}~\bibnamefont {Kulin}}, \bibinfo
  {author} {\bibfnamefont {R.}~\bibnamefont {Dumke}}, \bibinfo {author}
  {\bibfnamefont {S.~D.}\ \bibnamefont {Bergeson}}, \ and\ \bibinfo {author}
  {\bibfnamefont {S.~L.}\ \bibnamefont {Rolston}},\ }\href@noop {} {\bibfield
  {journal} {\bibinfo  {journal} {{ Phys. Rev. Lett.}}\ }\textbf {\bibinfo
  {volume} {86}},\ \bibinfo {pages} {3759} (\bibinfo {year}
  {2001})}\BibitemShut {NoStop}%
\bibitem [{\citenamefont {Fletcher}\ \emph {et~al.}(2007)\citenamefont
  {Fletcher}, \citenamefont {Zhang},\ and\ \citenamefont {Rolston}}]{fzr07}%
  \BibitemOpen
  \bibfield  {author} {\bibinfo {author} {\bibfnamefont {R.~S.}\ \bibnamefont
  {Fletcher}}, \bibinfo {author} {\bibfnamefont {X.~L.}\ \bibnamefont {Zhang}},
  \ and\ \bibinfo {author} {\bibfnamefont {S.~L.}\ \bibnamefont {Rolston}},\
  }\href {\doibase 10.1103/PhysRevLett.99.145001} {\bibfield  {journal}
  {\bibinfo  {journal} {Phys. Rev. Lett.}\ }\textbf {\bibinfo {volume} {99}},\
  \bibinfo {pages} {145001} (\bibinfo {year} {2007})}\BibitemShut {NoStop}%
\bibitem [{\citenamefont {Zhang}\ \emph
  {et~al.}(2008{\natexlab{a}})\citenamefont {Zhang}, \citenamefont {Fletcher},\
  and\ \citenamefont {Rolston}}]{zfr08instability}%
  \BibitemOpen
  \bibfield  {author} {\bibinfo {author} {\bibfnamefont {X.~L.}\ \bibnamefont
  {Zhang}}, \bibinfo {author} {\bibfnamefont {R.~S.}\ \bibnamefont {Fletcher}},
  \ and\ \bibinfo {author} {\bibfnamefont {S.~L.}\ \bibnamefont {Rolston}},\
  }\href {\doibase 10.1103/PhysRevLett.101.195002} {\bibfield  {journal}
  {\bibinfo  {journal} {Phys. Rev. Lett.}\ }\textbf {\bibinfo {volume} {101}},\
  \bibinfo {pages} {195002} (\bibinfo {year} {2008}{\natexlab{a}})}\BibitemShut
  {NoStop}%
\bibitem [{\citenamefont {Murillo}(2001)}]{mur01}%
  \BibitemOpen
  \bibfield  {author} {\bibinfo {author} {\bibfnamefont {M.~S.}\ \bibnamefont
  {Murillo}},\ }\href@noop {} {\bibfield  {journal} {\bibinfo  {journal}
  {{Phys. Rev. Lett.}}\ }\textbf {\bibinfo {volume} {87}},\ \bibinfo {pages}
  {115003} (\bibinfo {year} {2001})}\BibitemShut {NoStop}%
\bibitem [{\citenamefont {Bannasch}\ \emph {et~al.}(2012)\citenamefont
  {Bannasch}, \citenamefont {Castro}, \citenamefont {McQuillen}, \citenamefont
  {Pohl},\ and\ \citenamefont {Killian}}]{bcm12}%
  \BibitemOpen
  \bibfield  {author} {\bibinfo {author} {\bibfnamefont {G.}~\bibnamefont
  {Bannasch}}, \bibinfo {author} {\bibfnamefont {J.}~\bibnamefont {Castro}},
  \bibinfo {author} {\bibfnamefont {P.}~\bibnamefont {McQuillen}}, \bibinfo
  {author} {\bibfnamefont {T.}~\bibnamefont {Pohl}}, \ and\ \bibinfo {author}
  {\bibfnamefont {T.~C.}\ \bibnamefont {Killian}},\ }\href {\doibase
  10.1103/PhysRevLett.109.185008} {\bibfield  {journal} {\bibinfo  {journal}
  {Phys. Rev. Lett.}\ }\textbf {\bibinfo {volume} {109}},\ \bibinfo {pages}
  {185008} (\bibinfo {year} {2012})}\BibitemShut {NoStop}%
\bibitem [{\citenamefont {Sprenkle}\ \emph {et~al.}(2019)\citenamefont
  {Sprenkle}, \citenamefont {Dodson}, \citenamefont {McKnight}, \citenamefont
  {Spencer}, \citenamefont {Bergeson}, \citenamefont {Diaw},\ and\
  \citenamefont {Murillo}}]{sdm19}%
  \BibitemOpen
  \bibfield  {author} {\bibinfo {author} {\bibfnamefont {T.}~\bibnamefont
  {Sprenkle}}, \bibinfo {author} {\bibfnamefont {A.}~\bibnamefont {Dodson}},
  \bibinfo {author} {\bibfnamefont {E.}~\bibnamefont {McKnight}}, \bibinfo
  {author} {\bibfnamefont {R.}~\bibnamefont {Spencer}}, \bibinfo {author}
  {\bibfnamefont {S.}~\bibnamefont {Bergeson}}, \bibinfo {author}
  {\bibfnamefont {A.}~\bibnamefont {Diaw}}, \ and\ \bibinfo {author}
  {\bibfnamefont {M.~S.}\ \bibnamefont {Murillo}},\ }\href {\doibase
  10.1103/PhysRevE.99.053206} {\bibfield  {journal} {\bibinfo  {journal} {Phys.
  Rev. E}\ }\textbf {\bibinfo {volume} {99}},\ \bibinfo {pages} {053206}
  (\bibinfo {year} {2019})}\BibitemShut {NoStop}%
\bibitem [{\citenamefont {Robicheaux}\ and\ \citenamefont
  {Hanson}(2003)}]{rha03}%
  \BibitemOpen
  \bibfield  {author} {\bibinfo {author} {\bibfnamefont {F.}~\bibnamefont
  {Robicheaux}}\ and\ \bibinfo {author} {\bibfnamefont {J.~D.}\ \bibnamefont
  {Hanson}},\ }\href@noop {} {\bibfield  {journal} {\bibinfo  {journal} {{
  Phys. Plasmas}}\ }\textbf {\bibinfo {volume} {10}},\ \bibinfo {pages} {2217}
  (\bibinfo {year} {2003})}\BibitemShut {NoStop}%
\bibitem [{\citenamefont {Pohl}\ \emph
  {et~al.}(2004{\natexlab{a}})\citenamefont {Pohl}, \citenamefont {Pattard},\
  and\ \citenamefont {Rost}}]{ppr04PRA}%
  \BibitemOpen
  \bibfield  {author} {\bibinfo {author} {\bibfnamefont {T.}~\bibnamefont
  {Pohl}}, \bibinfo {author} {\bibfnamefont {T.}~\bibnamefont {Pattard}}, \
  and\ \bibinfo {author} {\bibfnamefont {J.~M.}\ \bibnamefont {Rost}},\
  }\href@noop {} {\bibfield  {journal} {\bibinfo  {journal} {{ Phys. Rev. A}}\
  }\textbf {\bibinfo {volume} {70}},\ \bibinfo {pages} {033416} (\bibinfo
  {year} {2004}{\natexlab{a}})}\BibitemShut {NoStop}%
\bibitem [{\citenamefont {Pohl}\ \emph
  {et~al.}(2004{\natexlab{b}})\citenamefont {Pohl}, \citenamefont {Pattard},\
  and\ \citenamefont {Rost}}]{ppr04PRL}%
  \BibitemOpen
  \bibfield  {author} {\bibinfo {author} {\bibfnamefont {T.}~\bibnamefont
  {Pohl}}, \bibinfo {author} {\bibfnamefont {T.}~\bibnamefont {Pattard}}, \
  and\ \bibinfo {author} {\bibfnamefont {J.~M.}\ \bibnamefont {Rost}},\ }\href
  {\doibase 10.1103/PhysRevLett.92.155003} {\bibfield  {journal} {\bibinfo
  {journal} {Phys. Rev. Lett.}\ }\textbf {\bibinfo {volume} {92}},\ \bibinfo
  {pages} {155003} (\bibinfo {year} {2004}{\natexlab{b}})}\BibitemShut
  {NoStop}%
\bibitem [{\citenamefont {Cummings}\ \emph {et~al.}(2005)\citenamefont
  {Cummings}, \citenamefont {Daily}, \citenamefont {Durfee},\ and\
  \citenamefont {Bergeson}}]{cdd05}%
  \BibitemOpen
  \bibfield  {author} {\bibinfo {author} {\bibfnamefont {E.~A.}\ \bibnamefont
  {Cummings}}, \bibinfo {author} {\bibfnamefont {J.~E.}\ \bibnamefont {Daily}},
  \bibinfo {author} {\bibfnamefont {D.~S.}\ \bibnamefont {Durfee}}, \ and\
  \bibinfo {author} {\bibfnamefont {S.~D.}\ \bibnamefont {Bergeson}},\
  }\href@noop {} {\bibfield  {journal} {\bibinfo  {journal} {Phys. Plasmas}\
  }\textbf {\bibinfo {volume} {12}},\ \bibinfo {pages} {123501} (\bibinfo
  {year} {2005})}\BibitemShut {NoStop}%
\bibitem [{\citenamefont {Gupta}\ \emph {et~al.}(2007)\citenamefont {Gupta},
  \citenamefont {Laha}, \citenamefont {Simien}, \citenamefont {Gao},
  \citenamefont {Castro}, \citenamefont {Killian},\ and\ \citenamefont
  {Pohl}}]{gls07}%
  \BibitemOpen
  \bibfield  {author} {\bibinfo {author} {\bibfnamefont {P.}~\bibnamefont
  {Gupta}}, \bibinfo {author} {\bibfnamefont {S.}~\bibnamefont {Laha}},
  \bibinfo {author} {\bibfnamefont {C.~E.}\ \bibnamefont {Simien}}, \bibinfo
  {author} {\bibfnamefont {H.}~\bibnamefont {Gao}}, \bibinfo {author}
  {\bibfnamefont {J.}~\bibnamefont {Castro}}, \bibinfo {author} {\bibfnamefont
  {T.~C.}\ \bibnamefont {Killian}}, \ and\ \bibinfo {author} {\bibfnamefont
  {T.}~\bibnamefont {Pohl}},\ }\href@noop {} {\bibfield  {journal} {\bibinfo
  {journal} {{ Phys. Rev. Lett.}}\ }\textbf {\bibinfo {volume} {99}},\ \bibinfo
  {pages} {75005} (\bibinfo {year} {2007})}\BibitemShut {NoStop}%
\bibitem [{\citenamefont {Laha}\ \emph {et~al.}(2007)\citenamefont {Laha},
  \citenamefont {Gupta}, \citenamefont {Simien}, \citenamefont {Gao},
  \citenamefont {Castro},\ and\ \citenamefont {Killian}}]{lgs07}%
  \BibitemOpen
  \bibfield  {author} {\bibinfo {author} {\bibfnamefont {S.}~\bibnamefont
  {Laha}}, \bibinfo {author} {\bibfnamefont {P.}~\bibnamefont {Gupta}},
  \bibinfo {author} {\bibfnamefont {C.~E.}\ \bibnamefont {Simien}}, \bibinfo
  {author} {\bibfnamefont {H.}~\bibnamefont {Gao}}, \bibinfo {author}
  {\bibfnamefont {J.}~\bibnamefont {Castro}}, \ and\ \bibinfo {author}
  {\bibfnamefont {T.~C.}\ \bibnamefont {Killian}},\ }\href@noop {} {\bibfield
  {journal} {\bibinfo  {journal} {{Phys. Rev. Lett.}}\ }\textbf {\bibinfo
  {volume} {99}},\ \bibinfo {pages} {155001} (\bibinfo {year}
  {2007})}\BibitemShut {NoStop}%
\bibitem [{\citenamefont {McQuillen}\ \emph
  {et~al.}(2015{\natexlab{b}})\citenamefont {McQuillen}, \citenamefont
  {Strickler}, \citenamefont {Langin},\ and\ \citenamefont {Killian}}]{msl15}%
  \BibitemOpen
  \bibfield  {author} {\bibinfo {author} {\bibfnamefont {P.}~\bibnamefont
  {McQuillen}}, \bibinfo {author} {\bibfnamefont {T.}~\bibnamefont
  {Strickler}}, \bibinfo {author} {\bibfnamefont {T.}~\bibnamefont {Langin}}, \
  and\ \bibinfo {author} {\bibfnamefont {T.~C.}\ \bibnamefont {Killian}},\
  }\href@noop {} {\bibfield  {journal} {\bibinfo  {journal} {Phys. Plasmas}\
  }\textbf {\bibinfo {volume} {22}},\ \bibinfo {pages} {033513} (\bibinfo
  {year} {2015}{\natexlab{b}})}\BibitemShut {NoStop}%
\bibitem [{\citenamefont {Zhang}\ \emph
  {et~al.}(2008{\natexlab{b}})\citenamefont {Zhang}, \citenamefont {Fletcher},
  \citenamefont {Rolston}, \citenamefont {Guzdar},\ and\ \citenamefont
  {Swisdak}}]{zfr08Bfield}%
  \BibitemOpen
  \bibfield  {author} {\bibinfo {author} {\bibfnamefont {X.~L.}\ \bibnamefont
  {Zhang}}, \bibinfo {author} {\bibfnamefont {R.~S.}\ \bibnamefont {Fletcher}},
  \bibinfo {author} {\bibfnamefont {S.~L.}\ \bibnamefont {Rolston}}, \bibinfo
  {author} {\bibfnamefont {P.~N.}\ \bibnamefont {Guzdar}}, \ and\ \bibinfo
  {author} {\bibfnamefont {M.}~\bibnamefont {Swisdak}},\ }\href {\doibase
  10.1103/PhysRevLett.100.235002} {\bibfield  {journal} {\bibinfo  {journal}
  {Phys. Rev. Lett.}\ }\textbf {\bibinfo {volume} {100}},\ \bibinfo {pages}
  {235002} (\bibinfo {year} {2008}{\natexlab{b}})}\BibitemShut {NoStop}%
\bibitem [{\citenamefont {Sprenkle}\ \emph {et~al.}(2022)\citenamefont
  {Sprenkle}, \citenamefont {Bergeson}, \citenamefont {Silvestri},\ and\
  \citenamefont {Murillo}}]{sbs22}%
  \BibitemOpen
  \bibfield  {author} {\bibinfo {author} {\bibfnamefont {R.~T.}\ \bibnamefont
  {Sprenkle}}, \bibinfo {author} {\bibfnamefont {S.~D.}\ \bibnamefont
  {Bergeson}}, \bibinfo {author} {\bibfnamefont {L.~G.}\ \bibnamefont
  {Silvestri}}, \ and\ \bibinfo {author} {\bibfnamefont {M.~S.}\ \bibnamefont
  {Murillo}},\ }\href@noop {} {\bibfield  {journal} {\bibinfo  {journal}
  {{Phys. Rev. E}}\ }\textbf {\bibinfo {volume} {105}},\ \bibinfo {pages}
  {045201} (\bibinfo {year} {2022})}\BibitemShut {NoStop}%
\bibitem [{\citenamefont {Warrens}\ \emph {et~al.}(2021)\citenamefont
  {Warrens}, \citenamefont {Gorman}, \citenamefont {Bradshaw},\ and\
  \citenamefont {Killian}}]{wgb21}%
  \BibitemOpen
  \bibfield  {author} {\bibinfo {author} {\bibfnamefont {M.~K.}\ \bibnamefont
  {Warrens}}, \bibinfo {author} {\bibfnamefont {G.~M.}\ \bibnamefont {Gorman}},
  \bibinfo {author} {\bibfnamefont {S.~J.}\ \bibnamefont {Bradshaw}}, \ and\
  \bibinfo {author} {\bibfnamefont {T.~C.}\ \bibnamefont {Killian}},\
  }\href@noop {} {\bibfield  {journal} {\bibinfo  {journal} {{Phys. Plasmas}}\
  }\textbf {\bibinfo {volume} {28}},\ \bibinfo {pages} {022110} (\bibinfo
  {year} {2021})}\BibitemShut {NoStop}%
\bibitem [{\citenamefont {Gorman}\ \emph {et~al.}(2021)\citenamefont {Gorman},
  \citenamefont {Warrens}, \citenamefont {Bradshaw},\ and\ \citenamefont
  {Killian}}]{gwb21}%
  \BibitemOpen
  \bibfield  {author} {\bibinfo {author} {\bibfnamefont {G.~M.}\ \bibnamefont
  {Gorman}}, \bibinfo {author} {\bibfnamefont {M.~K.}\ \bibnamefont {Warrens}},
  \bibinfo {author} {\bibfnamefont {S.~J.}\ \bibnamefont {Bradshaw}}, \ and\
  \bibinfo {author} {\bibfnamefont {T.~C.}\ \bibnamefont {Killian}},\
  }\href@noop {} {\bibfield  {journal} {\bibinfo  {journal} {{Phys. Rev.
  Lett.}}\ }\textbf {\bibinfo {volume} {126}},\ \bibinfo {pages} {085002}
  (\bibinfo {year} {2021})}\BibitemShut {NoStop}%
\bibitem [{\citenamefont {Vikhrov}\ \emph {et~al.}(2021)\citenamefont
  {Vikhrov}, \citenamefont {Bronin}, \citenamefont {Zelener},\ and\
  \citenamefont {Zelener}}]{vbz21}%
  \BibitemOpen
  \bibfield  {author} {\bibinfo {author} {\bibfnamefont {E.}~\bibnamefont
  {Vikhrov}}, \bibinfo {author} {\bibfnamefont {S.~Y.}\ \bibnamefont {Bronin}},
  \bibinfo {author} {\bibfnamefont {B.~B.}\ \bibnamefont {Zelener}}, \ and\
  \bibinfo {author} {\bibfnamefont {B.~V.}\ \bibnamefont {Zelener}},\
  }\href@noop {} {\bibfield  {journal} {\bibinfo  {journal} {{Phys. Rev. E}}\
  }\textbf {\bibinfo {volume} {104}},\ \bibinfo {pages} {015212} (\bibinfo
  {year} {2021})}\BibitemShut {NoStop}%
\bibitem [{\citenamefont {Bronin}\ \emph {et~al.}(2023)\citenamefont {Bronin},
  \citenamefont {Vikhrov}, \citenamefont {Zelener},\ and\ \citenamefont
  {Zelener}}]{bvz23}%
  \BibitemOpen
  \bibfield  {author} {\bibinfo {author} {\bibfnamefont {S.~Y.}\ \bibnamefont
  {Bronin}}, \bibinfo {author} {\bibfnamefont {E.~V.}\ \bibnamefont {Vikhrov}},
  \bibinfo {author} {\bibfnamefont {B.~B.}\ \bibnamefont {Zelener}}, \ and\
  \bibinfo {author} {\bibfnamefont {B.~V.}\ \bibnamefont {Zelener}},\ }\href
  {\doibase 10.1103/PhysRevE.108.045209} {\bibfield  {journal} {\bibinfo
  {journal} {Phys. Rev. E}\ }\textbf {\bibinfo {volume} {108}},\ \bibinfo
  {pages} {045209} (\bibinfo {year} {2023})}\BibitemShut {NoStop}%
\bibitem [{\citenamefont {Shukla}\ \emph {et~al.}(2011)\citenamefont {Shukla},
  \citenamefont {Moslem}, \citenamefont {Duha},\ and\ \citenamefont
  {Mamun}}]{smd11}%
  \BibitemOpen
  \bibfield  {author} {\bibinfo {author} {\bibfnamefont {P.}~\bibnamefont
  {Shukla}}, \bibinfo {author} {\bibfnamefont {W.~M.}\ \bibnamefont {Moslem}},
  \bibinfo {author} {\bibfnamefont {S.~S.}\ \bibnamefont {Duha}}, \ and\
  \bibinfo {author} {\bibfnamefont {A.~A.}\ \bibnamefont {Mamun}},\ }\href@noop
  {} {\bibfield  {journal} {\bibinfo  {journal} {{Europhys. Lett}}\ }\textbf
  {\bibinfo {volume} {96}},\ \bibinfo {pages} {65002} (\bibinfo {year}
  {2011})}\BibitemShut {NoStop}%
\bibitem [{\citenamefont {Sack}\ and\ \citenamefont {Schamel}(1985)}]{ssc85}%
  \BibitemOpen
  \bibfield  {author} {\bibinfo {author} {\bibfnamefont {C.}~\bibnamefont
  {Sack}}\ and\ \bibinfo {author} {\bibfnamefont {H.}~\bibnamefont {Schamel}},\
  }\href {\doibase 10.1088/0741-3335/27/7/002} {\bibfield  {journal} {\bibinfo
  {journal} {Plasma Physics and Controlled Fusion}\ }\textbf {\bibinfo {volume}
  {27}},\ \bibinfo {pages} {717} (\bibinfo {year} {1985})}\BibitemShut
  {NoStop}%
\bibitem [{\citenamefont {Dharodi}\ and\ \citenamefont
  {Murillo}(2020)}]{dmu20}%
  \BibitemOpen
  \bibfield  {author} {\bibinfo {author} {\bibfnamefont {V.~S.}\ \bibnamefont
  {Dharodi}}\ and\ \bibinfo {author} {\bibfnamefont {M.~S.}\ \bibnamefont
  {Murillo}},\ }\href@noop {} {\bibfield  {journal} {\bibinfo  {journal} {Phys.
  Rev. E}\ }\textbf {\bibinfo {volume} {101}},\ \bibinfo {pages} {023207}
  (\bibinfo {year} {2020})}\BibitemShut {NoStop}%
\bibitem [{\citenamefont {Killian}\ \emph {et~al.}(2012)\citenamefont
  {Killian}, \citenamefont {McQuillen}, \citenamefont {O'Neil},\ and\
  \citenamefont {Castro}}]{kmo12}%
  \BibitemOpen
  \bibfield  {author} {\bibinfo {author} {\bibfnamefont {T.~C.}\ \bibnamefont
  {Killian}}, \bibinfo {author} {\bibfnamefont {P.}~\bibnamefont {McQuillen}},
  \bibinfo {author} {\bibfnamefont {T.~M.}\ \bibnamefont {O'Neil}}, \ and\
  \bibinfo {author} {\bibfnamefont {J.}~\bibnamefont {Castro}},\ }\href
  {\doibase https://doi.org/10.1063/1.3694654} {\bibfield  {journal} {\bibinfo
  {journal} {Phys. Plasmas}\ }\textbf {\bibinfo {volume} {19}},\ \bibinfo
  {pages} {055701} (\bibinfo {year} {2012})}\BibitemShut {NoStop}%
\bibitem [{\citenamefont {Morrison}\ and\ \citenamefont {Grant}(2015)}]{mgr15}%
  \BibitemOpen
  \bibfield  {author} {\bibinfo {author} {\bibfnamefont {J.~P.}\ \bibnamefont
  {Morrison}}\ and\ \bibinfo {author} {\bibfnamefont {E.~R.}\ \bibnamefont
  {Grant}},\ }\href {\doibase 10.1103/PhysRevA.91.023423} {\bibfield  {journal}
  {\bibinfo  {journal} {Phys. Rev. A}\ }\textbf {\bibinfo {volume} {91}},\
  \bibinfo {pages} {023423} (\bibinfo {year} {2015})}\BibitemShut {NoStop}%
\bibitem [{\citenamefont {Viray}\ \emph {et~al.}(2020)\citenamefont {Viray},
  \citenamefont {Miller},\ and\ \citenamefont {Raithel}}]{vmr20}%
  \BibitemOpen
  \bibfield  {author} {\bibinfo {author} {\bibfnamefont {M.~A.}\ \bibnamefont
  {Viray}}, \bibinfo {author} {\bibfnamefont {S.~A.}\ \bibnamefont {Miller}}, \
  and\ \bibinfo {author} {\bibfnamefont {G.}~\bibnamefont {Raithel}},\
  }\href@noop {} {\bibfield  {journal} {\bibinfo  {journal} {{Phys. Rev. A}}\
  }\textbf {\bibinfo {volume} {102}},\ \bibinfo {pages} {033303} (\bibinfo
  {year} {2020})}\BibitemShut {NoStop}%
\bibitem [{\citenamefont {Metcalf}\ and\ \citenamefont {van~der
  Straten}(1999)}]{mvs99}%
  \BibitemOpen
  \bibfield  {author} {\bibinfo {author} {\bibfnamefont {H.~J.}\ \bibnamefont
  {Metcalf}}\ and\ \bibinfo {author} {\bibfnamefont {P.}~\bibnamefont {van~der
  Straten}},\ }\href@noop {} {\emph {\bibinfo {title} {Laser Cooling and
  Trapping}}}\ (\bibinfo  {publisher} {Springer-Verlag},\ \bibinfo {address}
  {New York, New York},\ \bibinfo {year} {1999})\BibitemShut {NoStop}%
\bibitem [{\citenamefont {Cooper}\ \emph {et~al.}(2018)\citenamefont {Cooper},
  \citenamefont {Covey}, \citenamefont {Madjarov}, \citenamefont {Porsev},
  \citenamefont {Safronova},\ and\ \citenamefont {Endres}}]{ccm18}%
  \BibitemOpen
  \bibfield  {author} {\bibinfo {author} {\bibfnamefont {A.}~\bibnamefont
  {Cooper}}, \bibinfo {author} {\bibfnamefont {J.~P.}\ \bibnamefont {Covey}},
  \bibinfo {author} {\bibfnamefont {I.~S.}\ \bibnamefont {Madjarov}}, \bibinfo
  {author} {\bibfnamefont {S.~G.}\ \bibnamefont {Porsev}}, \bibinfo {author}
  {\bibfnamefont {M.~S.}\ \bibnamefont {Safronova}}, \ and\ \bibinfo {author}
  {\bibfnamefont {M.}~\bibnamefont {Endres}},\ }\href {\doibase
  10.1103/PhysRevX.8.041055} {\bibfield  {journal} {\bibinfo  {journal} {Phys.
  Rev. X}\ }\textbf {\bibinfo {volume} {8}},\ \bibinfo {pages} {041055}
  (\bibinfo {year} {2018})}\BibitemShut {NoStop}%
\bibitem [{\citenamefont {Xu}\ \emph {et~al.}(2003)\citenamefont {Xu},
  \citenamefont {Loftus}, \citenamefont {Hall}, \citenamefont {Gallagher},\
  and\ \citenamefont {Ye}}]{xlh03}%
  \BibitemOpen
  \bibfield  {author} {\bibinfo {author} {\bibfnamefont {X.}~\bibnamefont
  {Xu}}, \bibinfo {author} {\bibfnamefont {T.~H.}\ \bibnamefont {Loftus}},
  \bibinfo {author} {\bibfnamefont {J.~L.}\ \bibnamefont {Hall}}, \bibinfo
  {author} {\bibfnamefont {A.}~\bibnamefont {Gallagher}}, \ and\ \bibinfo
  {author} {\bibfnamefont {J.}~\bibnamefont {Ye}},\ }\href {\doibase
  10.1364/JOSAB.20.000968} {\bibfield  {journal} {\bibinfo  {journal} {J. Opt.
  Soc. Am. B}\ }\textbf {\bibinfo {volume} {20}},\ \bibinfo {pages} {968}
  (\bibinfo {year} {2003})}\BibitemShut {NoStop}%
\bibitem [{\citenamefont {Nagel}\ \emph {et~al.}(2003)\citenamefont {Nagel},
  \citenamefont {Simien}, \citenamefont {Laha}, \citenamefont {Gupta},
  \citenamefont {Ashoka},\ and\ \citenamefont {Killian}}]{nsl03}%
  \BibitemOpen
  \bibfield  {author} {\bibinfo {author} {\bibfnamefont {S.~B.}\ \bibnamefont
  {Nagel}}, \bibinfo {author} {\bibfnamefont {C.~E.}\ \bibnamefont {Simien}},
  \bibinfo {author} {\bibfnamefont {S.}~\bibnamefont {Laha}}, \bibinfo {author}
  {\bibfnamefont {P.}~\bibnamefont {Gupta}}, \bibinfo {author} {\bibfnamefont
  {V.~S.}\ \bibnamefont {Ashoka}}, \ and\ \bibinfo {author} {\bibfnamefont
  {T.~C.}\ \bibnamefont {Killian}},\ }\href@noop {} {\bibfield  {journal}
  {\bibinfo  {journal} {{ Phys. Rev. A}}\ }\textbf {\bibinfo {volume} {67}},\
  \bibinfo {eid} {011401(R)} (\bibinfo {year} {2003})}\BibitemShut {NoStop}%
\bibitem [{\citenamefont {Chen}\ \emph {et~al.}(2004)\citenamefont {Chen},
  \citenamefont {Simien}, \citenamefont {Laha}, \citenamefont {Gupta},
  \citenamefont {Martinez}, \citenamefont {Mickelson}, \citenamefont {Nagel},\
  and\ \citenamefont {Killian}}]{csl04}%
  \BibitemOpen
  \bibfield  {author} {\bibinfo {author} {\bibfnamefont {Y.~C.}\ \bibnamefont
  {Chen}}, \bibinfo {author} {\bibfnamefont {C.~E.}\ \bibnamefont {Simien}},
  \bibinfo {author} {\bibfnamefont {S.}~\bibnamefont {Laha}}, \bibinfo {author}
  {\bibfnamefont {P.}~\bibnamefont {Gupta}}, \bibinfo {author} {\bibfnamefont
  {Y.~N.}\ \bibnamefont {Martinez}}, \bibinfo {author} {\bibfnamefont {P.~G.}\
  \bibnamefont {Mickelson}}, \bibinfo {author} {\bibfnamefont {S.~B.}\
  \bibnamefont {Nagel}}, \ and\ \bibinfo {author} {\bibfnamefont {T.~C.}\
  \bibnamefont {Killian}},\ }\href@noop {} {\bibfield  {journal} {\bibinfo
  {journal} {{Phys. Rev. Lett.}}\ }\textbf {\bibinfo {volume} {93}},\ \bibinfo
  {pages} {265003} (\bibinfo {year} {2004})}\BibitemShut {NoStop}%
\bibitem [{\citenamefont {Langin}\ \emph {et~al.}(2016)\citenamefont {Langin},
  \citenamefont {Strickler}, \citenamefont {Maksimovic}, \citenamefont
  {McQuillen}, \citenamefont {Pohl}, \citenamefont {Vrinceanu},\ and\
  \citenamefont {Killian}}]{lsm16}%
  \BibitemOpen
  \bibfield  {author} {\bibinfo {author} {\bibfnamefont {T.~K.}\ \bibnamefont
  {Langin}}, \bibinfo {author} {\bibfnamefont {T.}~\bibnamefont {Strickler}},
  \bibinfo {author} {\bibfnamefont {N.}~\bibnamefont {Maksimovic}}, \bibinfo
  {author} {\bibfnamefont {P.}~\bibnamefont {McQuillen}}, \bibinfo {author}
  {\bibfnamefont {T.}~\bibnamefont {Pohl}}, \bibinfo {author} {\bibfnamefont
  {D.}~\bibnamefont {Vrinceanu}}, \ and\ \bibinfo {author} {\bibfnamefont
  {T.~C.}\ \bibnamefont {Killian}},\ }\href@noop {} {\bibfield  {journal}
  {\bibinfo  {journal} {Phys. Rev. E}\ }\textbf {\bibinfo {volume} {93}},\
  \bibinfo {pages} {023201} (\bibinfo {year} {2016})}\BibitemShut {NoStop}%
\bibitem [{\citenamefont {Castro}\ \emph {et~al.}(2008)\citenamefont {Castro},
  \citenamefont {Gao},\ and\ \citenamefont {Killian}}]{cgk08}%
  \BibitemOpen
  \bibfield  {author} {\bibinfo {author} {\bibfnamefont {J.}~\bibnamefont
  {Castro}}, \bibinfo {author} {\bibfnamefont {H.}~\bibnamefont {Gao}}, \ and\
  \bibinfo {author} {\bibfnamefont {T.~C.}\ \bibnamefont {Killian}},\
  }\href@noop {} {\bibfield  {journal} {\bibinfo  {journal} {Plasma Phys.
  Control. Fusion}\ }\textbf {\bibinfo {volume} {50}},\ \bibinfo {pages}
  {124011} (\bibinfo {year} {2008})}\BibitemShut {NoStop}%
\bibitem [{\citenamefont {Szypko}\ \emph {et~al.}(2023)\citenamefont {Szypko},
  \citenamefont {Gorman},\ and\ \citenamefont {Bradshaw}}]{Szypko2023Modeling}%
  \BibitemOpen
  \bibfield  {author} {\bibinfo {author} {\bibfnamefont {G.}~\bibnamefont
  {Szypko}}, \bibinfo {author} {\bibfnamefont {G.}~\bibnamefont {Gorman}}, \
  and\ \bibinfo {author} {\bibfnamefont {S.}~\bibnamefont {Bradshaw}}\
  }(\bibinfo {year} {2023})\ \bibinfo {note}
  {https://baas.aas.org/pub/2023n7i107p04}\BibitemShut {NoStop}%
\bibitem [{\citenamefont {Centrella}\ and\ \citenamefont
  {Wilson}(1984)}]{Centrella1984}%
  \BibitemOpen
  \bibfield  {author} {\bibinfo {author} {\bibfnamefont {J.}~\bibnamefont
  {Centrella}}\ and\ \bibinfo {author} {\bibfnamefont {J.~W.}\ \bibnamefont
  {Wilson}},\ }\href@noop {} {\bibfield  {journal} {\bibinfo  {journal} {The
  Astrophysical Journal Suplement Series}\ }\textbf {\bibinfo {volume} {54}},\
  \bibinfo {pages} {229} (\bibinfo {year} {1984})}\BibitemShut {NoStop}%
\bibitem [{\citenamefont {Morawetz}\ and\ \citenamefont
  {R\"opke}(1996)}]{mro96}%
  \BibitemOpen
  \bibfield  {author} {\bibinfo {author} {\bibfnamefont {K.}~\bibnamefont
  {Morawetz}}\ and\ \bibinfo {author} {\bibfnamefont {G.}~\bibnamefont
  {R\"opke}},\ }\href {\doibase 10.1103/PhysRevE.54.4134} {\bibfield  {journal}
  {\bibinfo  {journal} {Phys. Rev. E}\ }\textbf {\bibinfo {volume} {54}},\
  \bibinfo {pages} {4134} (\bibinfo {year} {1996})}\BibitemShut {NoStop}%
\end{thebibliography}%
%

\end{document}